\documentclass{aa}  
\usepackage{aas_macros}

\usepackage{subcaption}
\usepackage{graphicx}

\usepackage{txfonts}
\usepackage{enumitem}

\usepackage{xcolor}
\usepackage{comment}
\usepackage[version=4]{mhchem}
\definecolor{burgundy}{rgb}{0.50,0.00,0.13}
\newcommand{\oh}[1]{\textnormal{#1}}
\usepackage{rotating} 

\newcommand{\ls}[1]{\textnormal{#1}}
\usepackage{sidecap}

\usepackage{scalerel}
\usepackage{tikz}
\usetikzlibrary{svg.path}
\usepackage{amssymb}

\definecolor{orcidlogocol}{HTML}{A6CE39}
\tikzset{
  orcidlogo/.pic={
    \fill[orcidlogocol] svg{M256,128c0,70.7-57.3,128-128,128C57.3,256,0,198.7,0,128C0,57.3,57.3,0,128,0C198.7,0,256,57.3,256,128z};
    \fill[white] svg{M86.3,186.2H70.9V79.1h15.4v48.4V186.2z}
                 svg{M108.9,79.1h41.6c39.6,0,57,28.3,57,53.6c0,27.5-21.5,53.6-56.8,53.6h-41.8V79.1z M124.3,172.4h24.5c34.9,0,42.9-26.5,42.9-39.7c0-21.5-13.7-39.7-43.7-39.7h-23.7V172.4z}
                 svg{M88.7,56.8c0,5.5-4.5,10.1-10.1,10.1c-5.6,0-10.1-4.6-10.1-10.1c0-5.6,4.5-10.1,10.1-10.1C84.2,46.7,88.7,51.3,88.7,56.8z};
  }
}

\usepackage{makecell}
\usepackage{ulem}
\usepackage{hyperref}
\begin{document} 

\title{SCoRE: the Surface Composition of Rocky Exoplanets}
\subtitle{Mineralogical constraints based on atmospheric types}

   \author{L. Sereinig 
            \inst{1, 2} 
            \and 
            O. Herbort 
            \inst{1} 
          }

   \institute{$^1$Institute for Astronomy (IfA), University of Vienna,
              T\"urkenschanzstrasse 17, A-1180 Vienna\\ 
              $^2$Now at: Centre for Planetary Habitability (PHAB), Department of Geosciences, University of Oslo, PO Box 1028 Blindern,
              0315 Oslo, Norway\\
              Contact: \email{leon.sereinig@geo.uio.no} \& \email{oliver.herbort@univie.ac.at}
             }

   \date{Received 04.03.2026; accepted 23.07.2026}

  \abstract
   {The crust composition of rocky exoplanets with a substantial atmosphere can not be observed directly. However, recent developments are starting to allow for the observation and characterisation of their atmospheres. Understanding the link between atmospheres and crusts could allow for constraints on the crusts' composition based on atmospheric observations.}  
   {We aim to understand the link between the thermal stability of specific condensates and atmospheric compositions. This allows constraints on the mineralogical composition of the surface by potentially observable atmospheric features. In order to achieve this, a grid study of atmospheric and crustal compositions is conducted. }
   {We use a diverse range of total element abundances inspired by various rock compositions as the compositional base for our crust-atmosphere models, for which thermo-chemical and phase equilibrium is assumed. In this work, we investigate a temperature range of 500\,K to 1000\,K and a surface pressure of 1\,bar. The resulting atmospheric compositions are classified using atmospheric types based on the most abundant gas-phase elements C, H, N, O, and S.}
   {Some of the changes in surface mineralogy coincide with changes in atmospheric type, independent of the given total elemental abundances. 
   In total, a link has been revealed between 23 thermally stable minerals and their corresponding atmospheric types, which is independent of the ratio of the refractory elements present. Especially, the sulphur chemistry of the minerals and the average iron oxidisation state can be constrained by the corresponding atmospheric type.}
   {}


   \maketitle
   \nolinenumbers
%
\section{Introduction}\label{sec:intro}

Over 8000 exoplanets have been detected at the time of writing\footnote{\url{https://exoplanet.eu/home/}, accessed March 2026}. Of those, about 12\% can be expected to be terrestrial exoplanets which posses a solid surface\footnote{\url{exoplanet.eu}, fraction of detected planets with $R_{\mathrm{p}}<1.5R_{\mathrm{Earth}}$}. Due to the smaller radii of rocky exoplanets compared to Jupiter- and sub-Neptune sized planets, terrestrial exoplanets are more difficult to detect, which leads to less frequent detections. The true population numbers of small, terrestrial exoplanets is currently unknown, and estimates of it might be biased towards larger Jupiter- or sub-Neptune sized exoplanets with close orbits around their host stars \citep[e.g.][]{Jiang24, bryson_earthestimate}.
Through various detection methods such as transit photometry and measurements of radial velocity variations, a range of parameters can be obtained. These are size, mass, and temperature \citep{Mayor-Queloz95, Henry00_size, Charbonneau_00_size}. The atmospheric composition can also be retrieved with e.g. transit spectroscopy \citep{Richardson_2007_spec, Pont_2008_spec}. Due to vast distances of exoplanets and small sizes of planets compared to the host star, the retrieval of these parameters from observations is challenging. For large, Jupiter-like planets it is possible to achieve good signal to noise ratios with current observational capabilities, but since terrestrial planets are multiple times smaller, observations are more difficult. Furthermore, terrestrial planets surfaces can starkly differ from one another (e.g. Venus', \oh{Earth's}, and Mars' surfaces). Methods such as polarimetry \citep{rossi_stam17}, or reflection and emission spectroscopy have been proposed as possible tools to infer crustal properties of exoplanets. 
In theory, it could be possible to constrain terrestrial exoplanets surfaces with reflection spectroscopy \citep[currently possible on solar system bodies, e.g.][]{Madden2018}. 
While these methods are potentially allowing for the characterisation of surfaces of bare rock planets, planets with substantial atmospheres are providing additional challenges. For example, reflection spectroscopy would have to limit itself to atmospheric windows, which in the extreme case of dense Venus-like atmospheres, can be narrow and few in numbers \citep{kappel2016multi}. Therefore, constraining the surface compositions of terrestrial exoplanets with a substantial atmosphere is impossible or difficult at best, making indirect approaches necessary.

Since the atmosphere and surface of rocky exoplanets are chemically interacting, their link provides an opportunity to constrain the planetary surface based on the observable atmosphere, which requires a good understanding of their connection. 
One approach to model this to a first degree, is to assume chemical and phase equilibrium,
\oh{which is likely to be reached for high temperatures. However, towards lower temperatures the timescales to reach equilibrium become increasingly longer. This applies to the gas phase \citep{Liggins2023} and the condensate phase \citep{Charnoz2026}.}
Since secondary atmospheres form largely from outgassing, which is governed by the planets interior \citep[see e.g.][]{spaargaren2020influence,baumeisterInterior, exogeo_spohn}, the bulk of these atmospheres form early on when the planet is still hot (leftover heat from planet formation). This means that equilibration time scales are short, and the system can therefore efficiently reach an equilibrium state between the atmosphere and the surface \citep[e.g.][]{Lichtenberg_2025}. Therefore, the assumption of a thermo-chemical and phase equilibrium state is reasonable for warm exoplanets. A similar approach has already been used for the modelling of Venus' surface by \cite{Byrne24}. 
This motivates that it is feasible to constrain the crustal composition of exoplanets by observing only their atmospheres. 

The observation of exoplanet atmospheric compositions is becoming increasingly more routine \citep[e.g.][]{guangwei_jwststatistics}, but still poses a challenge for small terrestrial worlds. Many attempted observations have returned non-detections \citep[e.g.][]{jwstDDT, ducrot2025_trap1b, piaulet2025strict_trap1d}. Recently, hot and possibly molten, terrestrial exoplanets have been observed to posses an atmosphere \citep{hu2024secondary, teske25_atmos}, but further observations are needed. Future missions and observatories such as the Extremely Large Telescope (ELT) \citep{gilmozzi2007european}, PLATO \citep{Rauer2025}, and Ariel \citep{Tinetti2018}, will provide the facilities needed to frequently measure atmospheric compositions of these small exoplanets. 

For the modelling of exoplanet crust-atmosphere pairs, elemental abundances should not only include volatiles such as CHNO (short for carbon, hydrogen, nitrogen, and oxygen), but also a range of refractory elements to accurately model the crust. For such sets of total elemental abundances, inspiration can be drawn from solar system bodies, such as Earth, CI chondrite \citep{loddersCI}, or asteroids that had samples returned to Earth such as Ryugu \citep[see][]{yokoyama2025elementalabundancesryuguassessment}. 

This work builds on the investigation of \cite{oh-ls25}, where an indication of an crust-atmosphere link has been reported. In this, a tentative link between the crustal compositions and atmospheric types based on the stability of molecules based on the elements C, H, and O has been observed. These atmospheric types have sharply defined transitions within a CHO-parameter space \citep[see Sect.~\ref{subsec:atmdiv} and ][]{Woitke21}, but because of the low number of crust-atmosphere models investigated therein, it remains unclear whether the changes in surface mineralogy are falling along the same transition lines as changes in the atmospheric types.
Therefore, this work will enhance this investigation by generating multiple model grids covering the entire CHO-atmospheric parameter space for different atmospheric temperatures, with each model in one grid based on the same set of total element abundances.
Since the range of exoplanet elemental compositions is currently unknown and likely very diverse, we use multiple different sets of total element abundances to generate multiple grids. The input compositions are inspired by solar system bodies and serve as starting points, attempting to reflect the possible diversity of atmospheres. We subsequently vary carbon, hydrogen, and oxygen abundances systematically per individual model in a grid, in order to archive different atmospheric compositions. The remaining mostly refractory elements are held constant. This enables us to generate a grid consisting of crust-atmosphere models spanning the entire CHO-atmospheric parameter space. By keeping the mostly refractory element abundances the same in every model, we isolate the effect of the atmospheric composition on the crustal composition. By using different base sets of relative elemental abundances for each grid, we rule out biases stemming from the input elemental abundance in the resulting crustal compositions.

The modelling of the crust-atmosphere pairs is achieved by calculation of the thermo-chemical and phase equilibrium. Due to the choice of the modelling tool, the model crusts comprise of only the pure form of condensates. In addition to these, solid solutions are also found in nature. However, modelling of these requires the usage of additional tools which is beyond this study. Here, we focus on the immediate interaction of the crust and atmosphere, which is why we only calculate the compositions of the crust and near-crust atmosphere \citep[introduced in][]{Herbort20rocky1}. 
The compositions of the near-crust atmospheres can be used as the basis for further atmospheric modelling, which is outside the scope of the investigation of the link between atmospheric types and the thermal stability of minerals.
An investigation of the link between specific clouds and surface conditions has been presented in \citet{Herbort2022}.
\citet{oh-ls25} have shown the potential differentiation of different atmospheric types by transmission spectroscopy. 

In this paper we investigate crustal compositions in phase equilibrium with atmospheric types A, B, C, $\alpha$, the transitional regime between types B and C/$\alpha$, and the sulphur-rich atmospheric type BC2. We define \ls{classes of condensate species depending on their stability and} characteristic crustal mineral compositions for each atmospheric type, independent of the total elemental abundance.

In Sect.~\ref{sec:methods} we provide an overview of our atmospheric and crust modelling methodology, give an overview of the atmospheric classification scheme in a CHO-parameter space, and give an explanation of how the model grids are generated. 
In Sect.~\ref{sec:results} links between certain crust condensates and their respective atmospheres are shown under the model assumptions. Section \ref{sec:limit} discusses the limitations caused by the assumption of thermo-chemical and phase equilibrium for the modelling of crust-atmosphere pairs, and discusses the implications of the the findings.
Section~\ref{sec:conclusion} provides a summary.

\section{Methods}
\label{sec:methods}

Crust-atmosphere models are generated using the equilibrium chemistry solver \textsc{GGchem} \citep{woitke2018equilibrium}. Using a constant pressure, a constant temperature, and a set of element abundances, the thermo-chemical and phase equilibrium is solved. The resulting composition consists of gas phase particles and condensate phase particles. The gas phase composition represents the atmospheric composition of the base layer of the atmosphere and the condensate phase represents the crust composition of the exoplanet. Atmospheric compositions using $p,T$-profiles which model a more complete atmosphere, have been previously investigated in \cite{oh-ls25}. The input set of element abundances consist of a set of 18 different elements, which are H, C, O, N, F, Na, Mg, Al, Si, P, S, Cl, K, Ca, Ti, Cr, Mn, and Fe. For these elements, \textsc{GGchem} includes a total of 471 gaseous species including molecules, atoms, and ions, and a total of 223 condensed species, of which 46 are liquid. 

\subsection{Atmospheric diversity}
\label{subsec:atmdiv}

In order to probe a large range of possible exoplanet atmospheres and their respective crust compositions multiple grids of crust-atmosphere models have been generated at temperatures ranging from 500\,K to 1000\,K, in 100\,K increments. The atmospheric pressure has been kept at a constant 1\,bar for all models. These model grids span a parameter space based on the three most abundant elements found in terrestrial exoplanets atmospheres, carbon (C), hydrogen (H), and oxygen (O), as described in \cite{Woitke21}. The CHO parameter space also defines atmospheric types, which are distinct regions where certain sets of molecular species coexist in thermo-chemical equilibrium. These regions define the atmospheric types as

\begin{align}
    \mathrm{Type~A:~} &\ce{H2O}, \ce{CH4}, \ce{NH3}, \ce{N2} \mathrm{~or~} \ce{H2};\nonumber \\
    \mathrm{Type~B:~} &\ce{H2O}, \ce{CO2}, \ce{N2}, \ce{O2};\label{eq:Types}\\
    \mathrm{Type~C:~} &\ce{H2O}, \ce{CO2}, \ce{CH4}, \ce{N2}; \nonumber\\
    \mathrm{Type~D:~} &\ce{CO2}, \ce{CH4}, \ce{CO} , \ce{N2}\nonumber.
\end{align}

\begin{figure}
    \centering
    \includegraphics[width=1\linewidth]{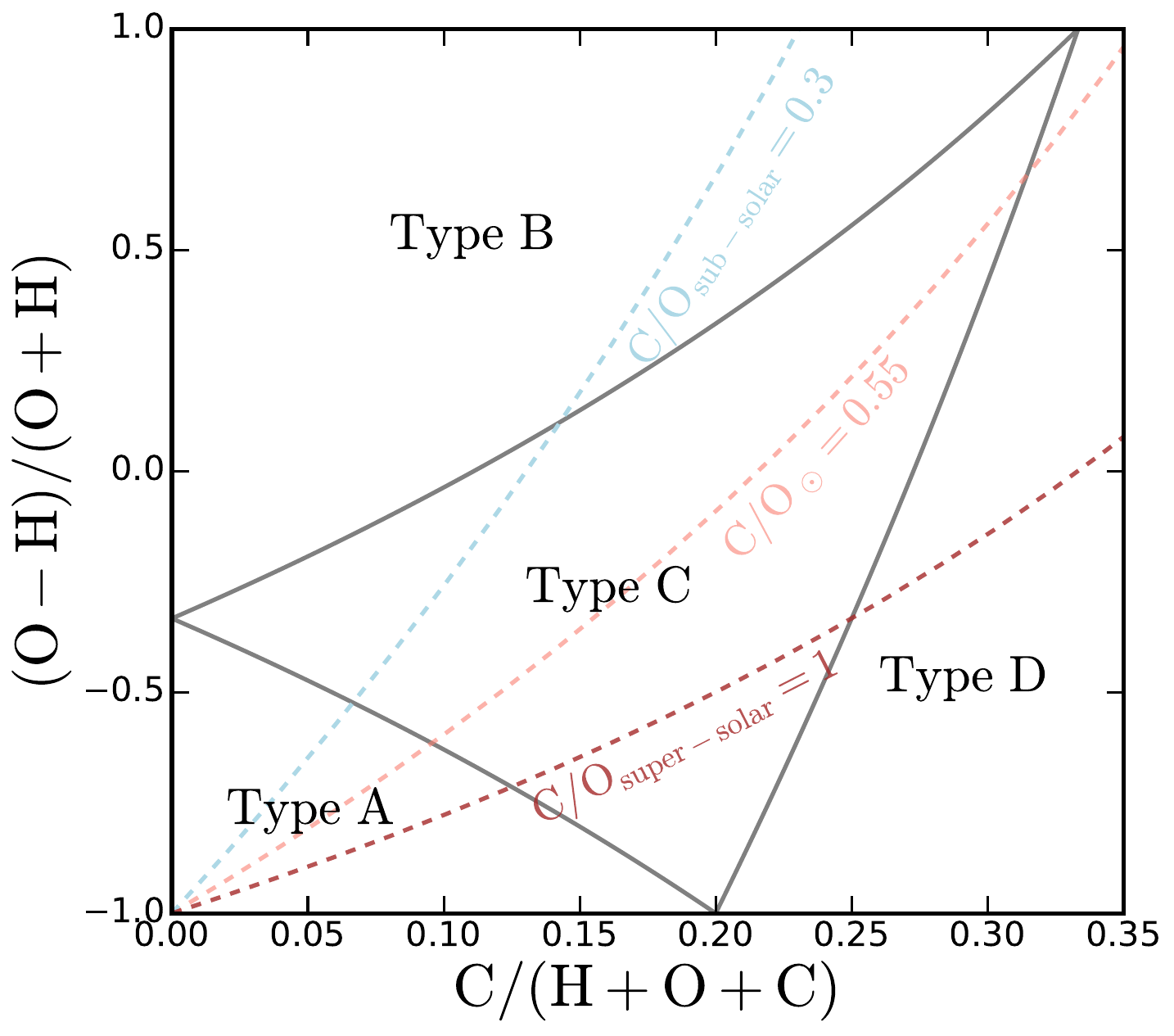}
    \caption{Atmospheric types as described by \cite{Woitke21}. Compositions of solar system bodies atmospheres are indicated. Dashed coloured lines show constant C/O-ratios for sub-solar, solar, and super-solar C/O ratios. }
    \label{fig:types600}
\end{figure}

The atmospheric parameter space is also displayed in Fig. \ref{fig:types600}. As stated in \cite{Woitke21}, these atmospheric types are distinct at temperatures up to 600\,K, beyond that types A, C, and D merge into one single type, indicated as type $\alpha$ (see Herbort \& Sereinig, in prep). Type B atmospheres remain unchanged at temperatures up to 1000\,K resulting in the following compositions in atmospheric types for temperatures of 700\,K to 1000\,K, with

\begin{align}
    \mathrm{Type~}\alpha\mathrm{~:~} &\ce{CO2}, \ce{H2O}, \ce{CH4}, \ce{CO}, \ce{H2}, \ce{N2};\nonumber \\
    \mathrm{Type~B:~} &\ce{O2}, \ce{CO2}, \ce{H2O}, \ce{N2}\label{eq:TypeshighT}.
\end{align}

\begin{figure}
    \centering
    \includegraphics[width=1\linewidth]{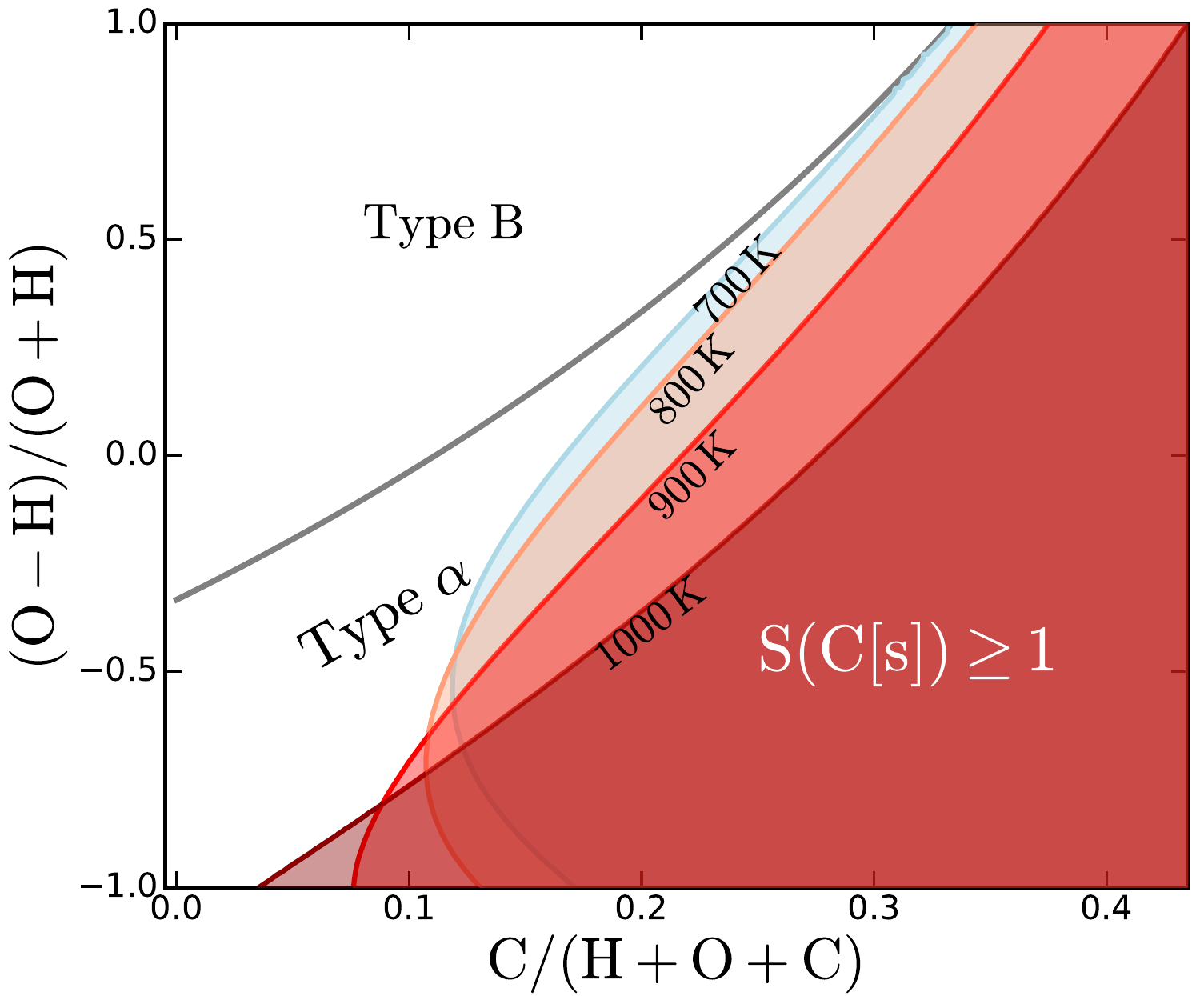}
    \caption{Atmospheric classification scheme for temperatures beyond 600\,K. Separate regimes at lower temperatures have merged into the homogeneous regime Type $\alpha$. Type B remains unaffected and has the same definition as in the low temperature case displayed in Fig. \ref{fig:types600}. The temperature evolution of graphite supersaturation is indicated with coloured surfaces. }
    \label{fig:types700}
\end{figure}

This evolved CHO-parameter space type classification for $T\geq700$\,K is displayed in Fig. \ref{fig:types700}.
If sufficient amounts of sulphur are abundant in the gas phase, compositions can lay beyond the CHO-parameter space where sulphur abundance has to be taken into account. \cite{Janssen_2023} expand the classification scheme from \cite{Woitke21} by the inclusion of sulphur. This leads to a number of new atmospheric subtypes, all of which contain sulphur compounds in the atmosphere. Most relevant for this work is the atmospheric type BC2, where \ce{SO2}, \ce{H2O}, \ce{CO2}, \ce{H2SO4}, and \ce{N2} coexist in thermo-chemical equilibrium in the gas phase. Atmospheres of this type can seemingly overlap with atmospheres of type B, if investigated in a CHO-parameter space. 

\subsection{Model grid generation}

In order to avoid biases from a chosen set of total element abundances, multiple input abundances based on different solar system bodies have been used for generating multiple crust-atmosphere model grids, with each grid based on a different set of total element abundances. Six sets of total element abundances (listed in table \ref{tab:abundances1}) have been used for calculating model grids at temperatures from 500\,K to 1000\,K, in 100\,K increments.

One set of total element abundances is an Earth inspired input abundance (Earth), taken from \cite{Herbort2022}, which under Earth like conditions (1.013\,bar, 288.15\,K) reproduces \oh{Earth's} atmospheric composition with equilibrium chemistry, \ls{with the only difference being a too low \ce{CH4} abundance.
However, \ce{CH4} is of biological origin and constantly produced. The co-existence of \ce{CH4} and \ce{O2} can be considered the disequilibrium signal in \oh{Earth's} atmosphere due to the presence of life.} 
Also inspired by Earth are Continental Crust (CC), Bulk Silicate Earth (BSE) \citep[both][]{schaefer2012}, and Mid Oceanic Ridge Basalt  \citep[MORB, a secondary crust composition in itself][]{arevaloMORB} set of elemental abundances. Inspired by the very rocks that form planets, asteroids, are CI Chondrite (CI) \citep{loddersCI}, and the composition of the asteroid Ryugu \cite{yokoyama2025elementalabundancesryuguassessment} sets of element abundances. Ryugus elemental abundances were measured by analysing material collected by the sample return mission Hayabusa-2. Since fluorine (F) abundance measurements were not performed on the returned samples, it was substituted by taking the value from CI and multiplying it by a factor $f$ defined as

\begin{align}
\label{eq:Fryugu}
    f= \frac{1}{17} \sum_{i=1}^{17} \frac{\epsilon_{\mathrm{Ryugu},i}}{\epsilon_{\mathrm{CI},i}},
\end{align}

where the $\epsilon_{\mathrm{abundance},i}$ are element abundances of Ryugu and CI, respectively. Since fluorine could not be included only the remaining 17 elements were used for this factor $f$. A complete list of the used input abundances are given in Appendix \ref{app:elemental-inputs}.

In order to generate models in a grid-like fashion, spanning the entire CHO atmospheric-parameter space based on one set of elemental abundances, the C, H, and O abundances of each model have been varied, while keeping the remaining elements constant.
This way, since most of the atmosphere consists of molecules formed from C, H, and O, the influence of the atmospheric composition on the crustal composition can be studied in isolation. Nitrogen has been kept constant, since it mainly forms \ce{N2}, and does not interfere with the majority of the atmospheric type system, as discussed in \cite{Woitke21}. Additionally, the ratio of volatile-to-refractory elements has been kept constant to the value of the volatile-to-refractory elemental ratio of each original set of elemental abundances. This was achieved by keeping the sum of $C+H+O$ constant while varying the individual C, H, and O abundances to generate the grid of crust-atmosphere models. Since other volatile elements such as N and S are kept constant the ratio of CHNOS-to-refractory elements remains constant for each individual model in a grid.

The set of total element abundances of MORB and BSE can not form full CHO parameter space spanning grids. This is because their low total $C+H+O$ budgets (44.542 and 44.32 mass fraction percent for MORB and BSE, respectively) do not allow for sufficient oxygen abundances to generate oxygen dominated type B atmospheres. Even if the entire  $C+H+O$ budget is spent on oxygen, the resulting atmospheric models are dominated by molecular nitrogen (\ce{N2}), with the defining molecular species for type B atmospheres only occurring in trace amounts. \oh{The numerically necessary residual H is sufficient to keep the atmospheres in type C.}

In total, four full parameter space spanning model grids per temperature have been generated for this work, based on the Earth, CC, CI chondrite, and Ryugu sets of total element abundances. Each grid consists of about 5000 crust-atmosphere models per temperature.

\section{Results}
\label{sec:results}
A total of 71 different condensates are stable across the model grids. For the individual temperature grids with 1000\,K, 900\,K, 800\,K, 700\,K, 600\,K, and 500\,K, a set of 34, 34, 48, 53, 48, and 61 condensates are thermally stable, respectively. 
For the interpretation of the results, we define four different classes  for the thermal stability of a condensate at a defined temperature. 

\begin{figure*}
    \centering
    \includegraphics[width=1\linewidth]{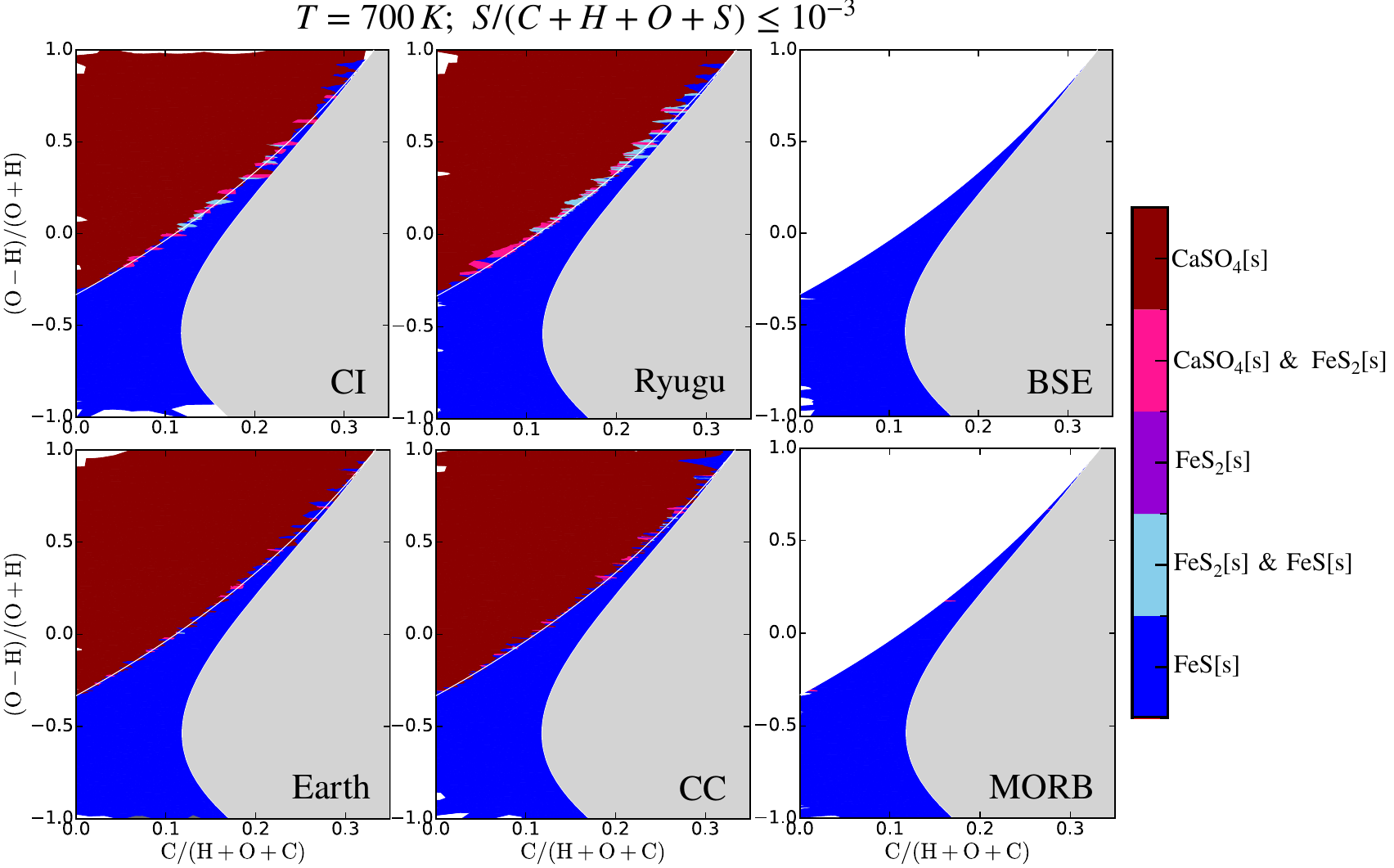}
    \caption{Solid sulphur in crusts at 700\,K for all set of total element abundances (indicated in each plot). Models where the atmosphere contains more than $S/(C+H+O+S)\geq10^{-3}$ are excluded. The transition from \ce{CaSO4}[s] in crusts in phase equilibrium with type B atmospheres, to \ce{FeS}[s] in crusts in phase equilibrium with type $\alpha$ atmospheres is clearly visible. All of the model grids based on different sets of total elemental abundances produce the same sequence of sulphur condensates throughout the atmospheric parameter space. This allows for the subsequent classification of the condensates displayed in this figure. Note that BSE and MORB elemental inputs do not allow for the generation of oxygen dominated type B atmospheres. Regions where the supersaturation ratio of carbon would exceed 1 are greyed out.}
    \label{fig:sulphurcond_poor}
\end{figure*}

\begin{itemize}[label={}, leftmargin=0pt, itemsep=3pt]
    \item{{\sffamily\large Class~0,~omnipresent:}} The condensate is present in all crusts, regardless of the contacting atmospheres type and set of total element abundances.
    \item{{\sffamily\large Class~I,~strongly~linked:}} The condensate is present in all crusts in phase equilibrium with atmospheres of a subset of the atmospheric parameter space (subset of atmospheric type, full atmospheric type, or multiple but not all atmospheric types), independent of the given set of total element abundances \ls{(see e.g. Fig. \ref{fig:sulphurcond_poor})}. 
    \item{{\sffamily\large Class~II,~partially~linked:}} The condensate is present in all crusts in phase equilibrium with atmospheres of a subset of the atmospheric parameter space, but can also exist in crusts in phase equilibrium with atmospheres in additional regimes of the CHO-parameter space, depending on the given set of total element abundances. 
    \item{{\sffamily\large Class~III,~unlinked:}} The condensate is unconstrained. Its stability likely strongly depends on the set of total element abundances of the model.
\end{itemize}

Each class is valid at a defined temperature, e.g. a Class 0 condensate at 600\,K can evolve into any of the other classes at different temperatures.
A comprehensive table including all condensates of Class 0-II, at every temperature, is provided in the appendix \ref{app:links}. A figure with representative crustal compositions and condensate classes is provided in the appendix in Fig. \ref{fig:TypeOverview}. Since this work is based on an equilibrium chemistry code which does not include solid solutions, each condensate is discussed as its own entity. However, in reality these condensates can exist in homogeneously mixed solid solutions.

Sulphur can not be classified as a fully volatile or refractory element. Depending on the pressure, temperature, and the availability of elements such as Ca, Fe, Si, and Al sulphur can be either part of the gas phase species or be thermally stable as a condensate. Therefore, the distinction between sulphur-poor atmospheres (Sect.~\ref{subsec:sulphurpoor_atm}) and sulphur-rich atmospheres (Sect.~\ref{subsec:sulphurrich_atm}) is made.

Sulphur containing condensates show a very stable link to the atmospheres composition across all investigated temperatures. This is because sulphur forms condensates with elements such as Fe and Ca which heavily depend on the atmospheric redox conditions. This link between sulphur bearing compounds and atmospheric composition is laid out in Sect. \ref{subsub:poor_sulphur_cond} for condensates under sulphur-poor atmospheres, and in Sect. \ref{subsub:rich_cond} for condensates in phase equilibrium with sulphur-rich atmospheres.

Iron condensates are inherently linked to the atmospheric composition due to iron's various oxidation states, which are intrinsically linked to the atmospheres redox potential. We note that pure iron (Fe[s]) is not directly linked to atmospheric type, since it requires extremely reducing atmospheric conditions to be stable as a condensate. The iron oxidation state of the surface and individual iron compounds links to the atmosphere are shown in Sects. \ref{subsub:iron_poor_cond} and \ref{subsub:rich_cond} for condensates in crusts in phase equilibrium with sulphur-poor and sulphur-rich atmospheres, respectively.

Two manganese species exhibit links to the atmospheric composition, which are described in Sect. \ref{subsub:mn_poor_cond} for crusts in phase equilibrium with sulphur-poor atmospheres, and in Sect.  \ref{subsub:rich_cond} for crusts in phase equilibrium with sulphur-rich atmospheres.

An additional 13 condensates, linked to the sulphur-poor atmospheric composition, are discussed in Sect. \ref{subsub:furth_poor_cond}.

\subsection{Condensates under sulphur-poor atmospheres}
\label{subsec:sulphurpoor_atm}
Sulphur-poor atmospheres can be described using the CHO-parameter space system, introduced by \cite{Woitke21}. Here we define the sulphur-poor atmospheres as such, that the sulphur content does not exceed  $S/(C+H+O+S)\leq0.005$ at 1000\,K, $S/(C+H+O+S)\leq 0.002$ at 900\,K and 800\,K, and $S/(C+H+O+S) \leq 0.001$ for $T\leq700$\,K. This choice is motivated by the fact that overall less condensates are thermally stable in chemical and phase equilibrium at higher temperatures, and sulphur is more abundant in the gas phase as a consequence.

\subsubsection{Sulphur containing condensates}
\label{subsub:poor_sulphur_cond}

We find three Class I sulphur bearing condensates.
These are \ce{CaSO4}[s] (calcium sulfate), \ce{FeS}[s] (iron(II) sulfide), and \ce{FeS2}[s] (pyrite). All are Class I condensates for $T\leq800$\,K. For temperatures $T>800$\,K, \ce{FeS2}[s] is not thermally stable \citep{jovanovic1989kinetics}. The transition between the different sulphur condensate species in crusts across the atmospheric parameter space is shown in Fig. \ref{fig:sulphurcond_poor} for $T=700$\,K.

Crusts in phase equilibrium with atmospheres of type A and C for $T\leq600$\,K, type $\alpha$ for $T\geq700$\,K, contain \ce{FeS}[s] as the only sulphur bearing compound (blue in Fig. \ref{fig:sulphurcond_poor}). However, two exceptions to this can be seen in the data. Firstly, in crusts in phase equilibrium with the most reducing atmospheres, where hydrogen and carbon abundances are high. There, iron is only available as Fe[s], preventing the formation of \ce{FeS}[s]. In those crusts, sulphur exists as MnS[s], MgS[s], or CaS[s] (also discussed in Sect. \ref{sub:outliers}). The second exception are models based on the Earth set of total element abundances at temperatures from 800\,K to 1000\,K. There, type $\alpha$ atmospheres tend to contain up to 0.18\% of $S/(C+H+O+S)$ in the gas phase, which leaves no sulphur available to condense (see also Fig. \ref{fig:summary}). This can be explained by the low sulphur abundance in the Earth based set of total elemental input abundances, as this does not happen in the similar CC set of total element abundances based grid, because there is enough sulphur present in the set of total element abundances for sulphur to be stable in both the gas and condensate phase.

Oxidising type B atmospheres give rise to crusts which contain sulphur exclusively as sulphates (\ce{-SO4}[s] compounds), most predominant as \ce{CaSO4}[s] (brown in Fig. \ref{fig:sulphurcond_poor}). Only in crusts based on the total elemental abundances of Ryugu and CI, \ce{CaSO4}[s] coexists with \ce{Na2SO4}[s]. This is due to the the disparity between the low abundance of calcium and high sulphur abundance in the elemental input of Ryugu and CI based models. Regardless, all crusts in phase equilibrium with type B atmospheres contain \ce{CaSO4}[s] at all temperatures investigated in this work.

\begin{figure}
    \centering
    \includegraphics[width=1\linewidth]{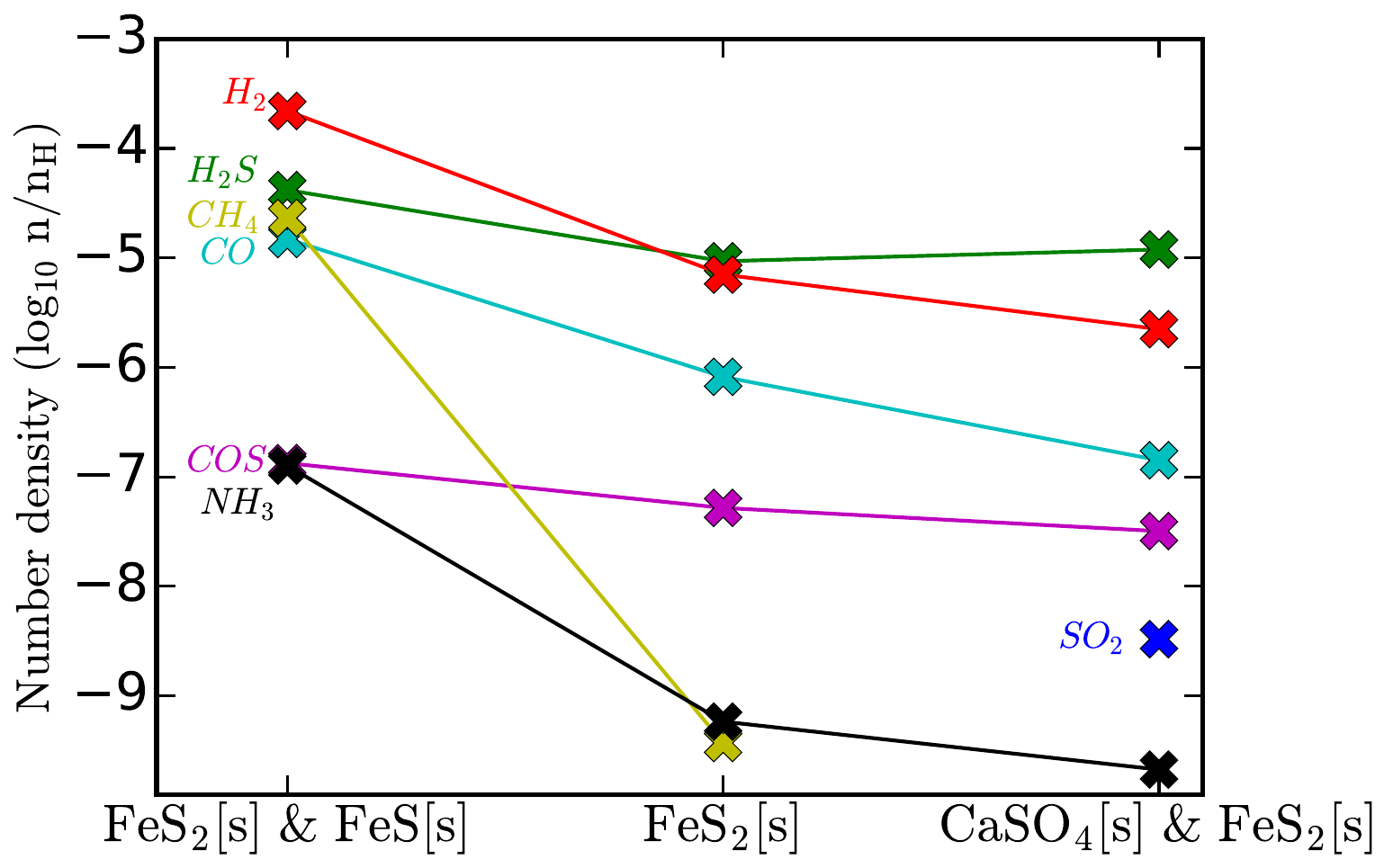}
    \caption{Atmospheric trace species number densities in the transitional regime. Reducing species become less abundant while oxidising become more abundant. Corresponding sulphur condensates are displayed on the x-axis. }
    \label{fig:typeT}
\end{figure}

In between the two reducing and oxidising regimes type C/$\alpha$ and type B, there is a narrow transitional region of slightly different atmospheric compositions. These atmospheres contain marginally more sulphur, in the form of sulphur containing trace species in the gas phase. Crusts in phase equilibrium with these atmospheres contain sulphur in three different forms, but always with the inclusion of \ce{FeS2}[s]. Under the most reducing environment, crusts contain sulphur as \ce{FeS}[s] in combination with \ce{FeS2}[s]. In the slightly more oxidising environment in this transitional regime of atmospheres, crusts contain \ce{FeS2}[s] exclusively. In crusts in phase equilibrium with the most oxidising compositions in this transitional regime, sulphur exists as \ce{FeS2}[s] in combination with [Ca, \ce{Na2}]\ce{SO4}[s] compounds. This transition in relation to gas phase species number densities is displayed in Fig. \ref{fig:typeT}. The different crust condensate compositions displayed in the atmospheric CHO-parameter space \ls{can be seen in Fig. \ref{fig:sulphurcond_poor} along the atmospheric type border between type alpha and B.}

Overall, sulphur condensates go from \ce{FeS}[s] in crusts in phase equilibrium with more reducing atmospheric compositions in type A and C at $T\leq600$\,K and type $\alpha$ at $T\geq700$\,K, to -\ce{SO4}[s] compounds in crusts in phase equilibrium with oxidising type B atmospheres. In between the two regimes lies a narrow transitional class of atmospheric and crustal compositions.

\subsubsection{Iron in crusts under sulphur-poor atmospheres}
\label{subsub:iron_poor_cond}

There are six linked iron containing compounds seen in the models. Iron's oxidation states Fe$^{2+}$ and Fe$^{3+}$ will be referred to as iron(II) and iron(III), respectively. \ce{Fe2O3}[s] (iron(III) oxide) is a Class I condensate at all temperatures examined in this work. Further linked condensate species include \ce{Fe3O4}[s] (iron(II,III) oxide), a Class I condensate at all temperatures examined in this work, except at 600\,K where it is classified as Class II. \ce{FeTiO3}[s] (ilmenite) is a Class II condensate at 500\,K and a Class I condensate for temperatures $T\geq600$\,K. \ce{FeSiO4}[s] (fayalite) is a Class~I condensate for $T\geq700$\,K. As \ce{FeS}[s] and \ce{FeS2}[s] also contain sulphur, they have been discussed in Sect. \ref{subsub:poor_sulphur_cond}. 

\ce{Fe2O3}[s], as an iron(III) compound, exists under oxidising conditions found in all crusts in phase equilibrium with type B atmospheres, at all temperatures investigated in this work. 

\ce{Fe3O4}[s] is an iron(II,III) compound, therefore it is stable under moderately oxidising environments. At 500\,K it exists in crusts in phase equilibrium with type C atmospheres, however depending on the set of total element abundances, it appears only in crusts in phase equilibrium with atmospheres in a complex area in the parameter space (see also upper left panel in Fig. \ref{fig:summary}). In model grids using the Earth and CC elemental abundance, \ce{Fe3O4}[s] is only stable in crusts in phase equilibrium with type C atmospheres with $C/(H+O+C)\geq0.1$ and along the type B-C border. At 600\,K in model grids based on Ryugu and CI sets of total element abundances it appears in all crusts in phase equilibrium with the transitional region between types C and B. 
For $T\geq700$\,K, \ce{Fe3O4}[s] exists in all crusts in phase equilibrium with atmospheres part of the transitional regime between type B and type $\alpha$ atmospheres as a Class I condensate.

\ce{FeTiO3}[s] is an iron (II) compound requiring reducing conditions. At 500\,K it exists in crusts in phase equilibrium with atmospheres in the transitional region between types B and C as a Class II condensate. Depending on the set of total element abundances, it can also exist in crusts in phase equilibrium with type A and C atmospheres. For $T\geq600$\,K it is a Class I condensate appearing in crusts in phase equilibrium with type A and C atmospheres at 600\,K, and type $\alpha$ for $T\geq700$\,K.

\begin{figure}
    \centering
    \includegraphics[width=1\linewidth]{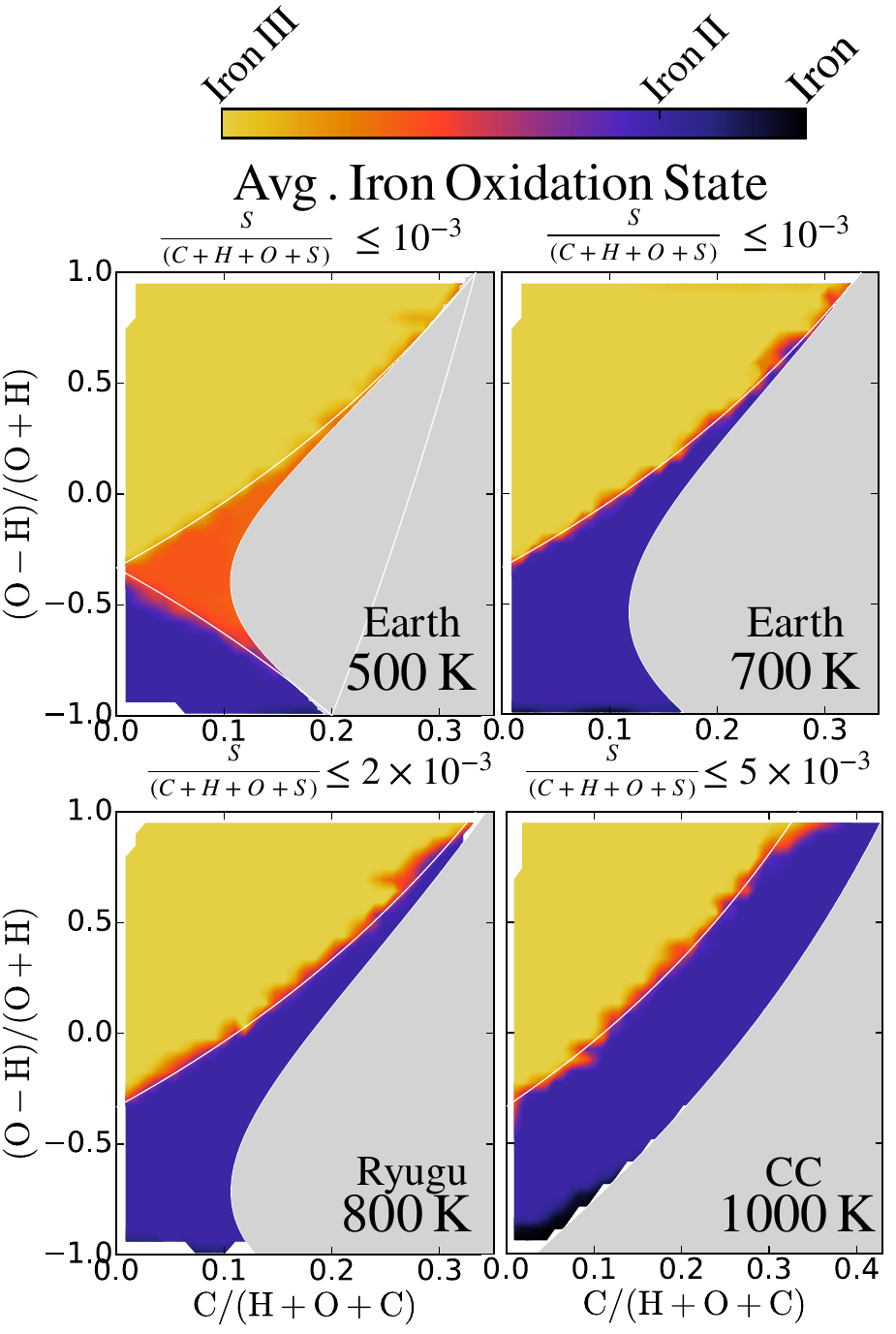}
    \caption{\ls{Iron oxidation state of crusts across the parameter space. Sulphur-rich models are excluded by the application of temperature dependent sulphur cut-off values as indicated above each panel (see Sect. \ref{subsec:sulphurpoor_atm}). The upper left panel shows the split in the crusts iron oxidation state between type A and C, which is dependent on the set of total elemental abundances}. The upper right and lower left panels display the `nominal' case, with a thin transitional region between the atmospheric types C and B. The lower right panel shows the high temperature case with an expanded region of stability for pure unoxidised iron in highly reducing environments.}
    \label{fig:iron_poor}
\end{figure}

\ce{Fe2SiO4}[s] is a Class II condensate at 500\,K, where it exists in crusts in phase equilibrium with a subset of type A atmospheres (for the exact region see Fig. \ref{fig:summary}). In the model grid based on the CC set of total element abundances at 500\,K, it also occurs in crusts where C[s] (graphite) is stable. At 600\,K, \ce{Fe2SiO4}[s] is also a Class II condensate appearing in crusts in phase equilibrium with type A atmospheres, with the Earth and CC abundance-based model grids also yielding  \ce{Fe2SiO4}[s] in crusts in phase equilibrium with type C atmospheres. From 700\,K to 1000\,K it is a Class I condensate, appearing in all crusts in phase equilibrium with type $\alpha$ atmospheres.

Based on all iron containing condensates, the average oxidation state is calculated, shown in Fig. \ref{fig:iron_poor}. The link between oxidation state and atmospheric type is clearly visible.

Model grids generated at 500\,K show that crusts in phase equilibrium with type A atmospheres have the average iron oxidation state of slightly below iron(II), with the iron in crusts in phase equilibrium with the most reducing atmospheric compositions (high H and C abundance) being pure and unoxidised iron. Crusts in phase equilibrium with type C atmospheres have an average iron oxidation state of slightly above iron(II) or iron(II,III), depending on input composition. This is shown in Fig \ref{fig:iron_poor} (upper left panel). Iron rich compositions (CI, Ryugu) tend to have crusts with lower average iron oxidation states. Crusts in phase equilibrium with type B atmospheres have an average oxidation state of iron(III) exclusively.

At 600\,K, crusts in phase equilibrium with type A atmospheres, have the average iron oxidation state of slightly below iron(II), with crusts in the most reducing environment (high H and C) containing pure unoxidised iron. Crusts in phase equilibrium with type C atmospheres have the average iron oxidation state of slightly above iron(II). Crusts under oxidising environments such as type B atmospheres contain iron in the state of iron(III) exclusively.

From 700\,K to 1000\,K, there is little evolution in the crusts average iron oxidation state. At 700\,K, crusts in phase equilibrium with type $\alpha$ generally adopt an average iron oxidation state of about iron(II) for all temperatures up to 1000\,K. With rising temperatures, the region where pure unoxidised iron is stable expands slightly, which is displayed the lower right panel in Fig. \ref{fig:iron_poor}. Crusts in phase equilibrium with type B atmospheres have the average iron oxidation state of iron(III) from 700\,K to 1000\,K.

Figure \ref{fig:iron_poor} also shows that the assumption of a constant $O/H$ or $(O-H)/(O+H)$ ratio is not sufficient to estimate the crusts average iron oxidation state. For any given $(O-H)/(O+H)$ ratio at $T\leq500$\,K, crusts can be in phase equilibrium with atmospheres of either type, depending on the carbon content of the atmosphere. For $T\geq700$\,K, atmospheres with $(O-H)/(O+H) > -\frac{1}{3}$ (corresponding to $O/H>0.5$) can harbour crusts with differing average iron oxidation states depending on the corresponding atmospheres type. Since the crusts average oxidation state depends on the contacting atmospheres type, a prediction of it based solely on the $(O-H)/(O+H)$ ratio is therefore not possible.

\subsubsection{Manganese containing species under sulphur-poor atmospheres}
\label{subsub:mn_poor_cond}

The manganese condensate species \ce{Mn3Al2Si3O12}[s] (spessartine) and \ce{Mn2O3}[s] (manganese(III) oxide) are both linked to atmospheric type.  

At 500\,K \ce{Mn3Al2Si3O12}[s] is a Class II condensate, found in crusts in phase equilibrium with atmospheres in the transitional regime between types C and B. Furthermore, depending on the set of total element abundances given it can also appear in crusts in phase equilibrium with types A and C. At 600\,K \ce{Mn3Al2Si3O12}[s] is a Class I condensate, appearing in all crusts in phase equilibrium with types A and C atmospheres. From 700\,K to 1000\,K it is also classified as Class I condensate for crusts in phase equilibrium with type $\alpha$ atmospheres.

\begin{figure}
    \centering
    \includegraphics[width=1\linewidth]{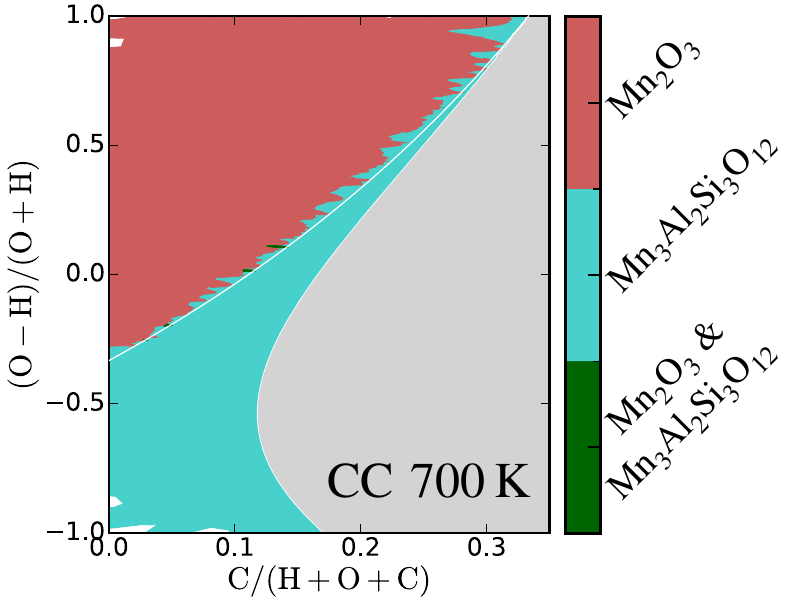}
    \caption{\ce{Mn3Al2Si3O12}[s] and \ce{Mn2O3}[s] in crusts in phase equilibrium with atmospheres across the CHO-parameter space.}
    \label{fig:Manganese}
\end{figure}

\ce{Mn2O3}[s] contains manganese in the form of manganese(III), the 3+ oxidation state of manganese, which requires oxidising environments found in crusts in phase equilibrium with type B atmospheres. At 500\,K it is a Class III condensate (unconstrained, because it does not appear in the model grid based on the Earth set of total element abundances), whereas at 600\,K it is a Class I condensate appearing in crusts in phase equilibrium with type B atmospheres above $(O-H)/(O+H)=0.3$. Below that, in crusts in phase equilibrium with atmospheres with $(O-H)/(O+H)\leq0.3$, it can be replaced with \ce{Ca2MnAl2Si3O13H}[s], depending on the set of total element abundances. For $T=700$\,K up to $T=1000$\,K, \ce{Mn2O3}[s] is a Class I condensate in crusts in phase equilibrium with type B atmospheres.

\ls{It is worth mentioning that in rare instances both manganese containing species appear together in crusts in phase equilibrium with atmospheres in the transitional region between types B and $\alpha$ (see also Fig. \ref{fig:Manganese}).} This bi-modal distribution of manganese containing condensates across the CHO-parameter space is shown in Fig. \ref{fig:Manganese}.

\subsubsection{Further condensate species under sulphur-poor atmospheres}
\label{subsub:furth_poor_cond}

\ce{MgSiO3}[s] (enstatite) is an unconstrained Class III condensate at 500\,K. At  600\,K and 700\,K it is a Class 0 condensate. For 800\,K to 1000\,K, \ce{MgSiO3}[s] is a Class II condensate existing in all crusts in phase equilibrium with type B atmospheres, with the Earth and CC set of total element abundances yielding models where \ce{MgSiO3}[s] is stable in crusts in phase equilibrium with type $\alpha$ atmospheres as well. Crusts at 800\,K and 900\,K, in phase equilibrium with type $\alpha$ atmospheres, based on the Ryugu and CI set of total element abundances only contain \ce{MgSiO3}[s] in a subset of type $\alpha$.

\ce{TiO2}[s] (titanium dioxide) is a Class I condensate at 500\,K, with the caveat that it is only a Class I condensate in crusts in phase equilibrium with atmospheres with $(O-H)/(O+H)\leq0.3$. In models based on the Earth and CC sets of total element abundances, crusts in phase equilibrium with type B atmospheres where the $(O-H)/(O+H)$ ratio is greater than 0.3, \ce{CaTiSiO5}[s] (titanite) takes its place. For $T>500$\,K it is a Class III (unconstrained) condensate.

\ce{CaTiSiO5}[s] (titanite) at 600\,K and 700\,K, \ce{CaTiSiO5}[s] exists in crusts in phase equilibrium with atmospheres in the transitional regime between type C/$\alpha$ and B. From  800\,K to 1000\,K \ce{CaTiSiO5}[s] is a Class III unconstrained condensate.

\ce{SiO2}[s] (quartz) is a Class II condensate for $T>500$\,K. At 500\,K it is unconstrained, at 600\,K it exists in crusts in phase equilibrium with type B atmospheres. In models using the Earth and CC set of total element abundances \ce{SiO2}[s] also exists in crusts in phase equilibrium with type A and C atmospheres. From 700\,K to 1000\,K \ce{SiO2}[s] exists as a Class II condensate in crusts in phase equilibrium with type B atmospheres. In models based on the Earth and CC set of total element abundances \ce{SiO2}[s] also exists in crusts in phase equilibrium with type $\alpha$ atmospheres.

\ce{Cr2O3}[s] (chromium(III) oxide) appears as a Class II condensate at every temperature examined in this work. \ce{Cr2O3}[s] is stable in crusts in phase equilibrium with type B atmospheres. In model grids based on the Earth and CC set of total element abundances, \ce{Cr2O3}[s] also occurs in crusts in phase equilibrium with type A and C atmospheres at $T\leq600$\,K, and in crusts in phase equilibrium with type $\alpha$ atmospheres at $T>600$\,K atmospheres. 

\ce{CaMgSi2O6}[s] (diopside) is found in crusts in phase equilibrium with type A atmospheres at 500\,K. \ce{Ca5P3O12F}[s] (fluorapatite) is found in crusts in phase equilibrium with type A and C atmospheres at 500\,K. Above $T\geq600$\,K, \ce{CaMgSi2O6}[s] and \ce{Ca5P3O12F}[s] are Class II condensates and behave the same. They exist in crusts in phase equilibrium with type A and C atmospheres, and at $T\geq700$\,K, type $\alpha$ atmospheres. Both condensates can also occur in crusts in phase equilibrium with type B atmospheres in models based on the Earth and CC set of total element abundances. 

The feldspar endmembers \ce{KAlSi3O8}[s] (orthoclase), \ce{CaAl2Si2O8}[s] (anorthite), and \ce{NaAlSi3O8}[s] (albite) show various links to the atmosphere. \ce{KAlSi3O8}[s] is a Class II condensate in crusts in phase equilibrium with type B atmospheres at 500\,K, 600\,K and 700\,K, and a Class 0 condensate for $T \geq 800$\,K.
At 500\,K \ce{KAlSi3O8}[s] only exists in crust in phase equilibrium with the most oxidising atmospheric compositions with $(O-H)/(O+H)\gtrsim0.9$. \ce{NaAlSi3O8}[s] is a Class III condensate at 500\,K and 600\,K. \ce{CaAl2Si2O8}[s] is a Class III unconstrained condensate at 500\,K and a Class II condensate at 600\,K, occurring in crusts in phase equilibrium with type A and C atmospheres.\ce{CaAl2Si2O8}[s] and \ce{NaAlSi3O8}[s] behave the same for $T\geq700$\,K, both being Class II condensates in crusts in phase equilibrium with type $\alpha$ atmospheres up to 1000\,K. 

\ce{MgF2}[s] (magnesium flouride) is linked to the atmospheric type at 500\,K, where it exists in crusts in phase equilibrium with type B atmospheres. The grid based on the Earth set of total element abundances also yields \ce{MgF2}[s] in crusts in phase equilibrium with atmospheres below $-0.9\gtrsim(O-H)/(O+H)$ in type A. At 600\,K and 700\,K, \ce{MgF2}[s] is a Class II condensate in crusts in phase equilibrium with type B atmospheres.\ls{ At 800\,K, \ce{MgF2}[s] behaves the same as at lower temperatures in models, with the exception of the CI-based model grid. There, it only exists in models above $0.9\approx(O-H)/(O+H)$.} At 900\,K and 1000\,K \ce{MgF2}[s] stops appearing in CI  set of total element abundances based models altogether and is unconstrained.

\ce{Mg3Si4O12H2}[s] (talc) is a Class II condensate in crusts in phase equilibrium with type B atmospheres at 500\,K. Depending on the set of total element abundances it is also found in crusts in phase equilibrium with type C and A atmospheres. At all other temperatures investigated in this work it is a Class III condensate.

\ce{NaCl}[s] (sodium chloride) is a Class II condensate for 600\,K. At this temperature it is found in crusts in phase equilibrium with type A and C atmospheres. Additionally, in grids based on Earth and CC sets of total element abundances, \ce{NaCl}[s] is also present in crusts in phase equilibrium with type B atmospheres. 

A graphical representation of all of the discussed condensates (including this and previous sections), excluding condensates purely found in the transitional regime between types C and B, is provided in Fig. \ref{fig:summary}.\ls{ A tabulated version of these results is provided in Tab. \ref{tab:crustcond1} and \ref{tab:crustcond2}.}

\subsection{Condensates under sulphur-rich atmospheres}
\label{subsec:sulphurrich_atm}

\begin{figure}
    \centering
    \includegraphics[width=1\linewidth]{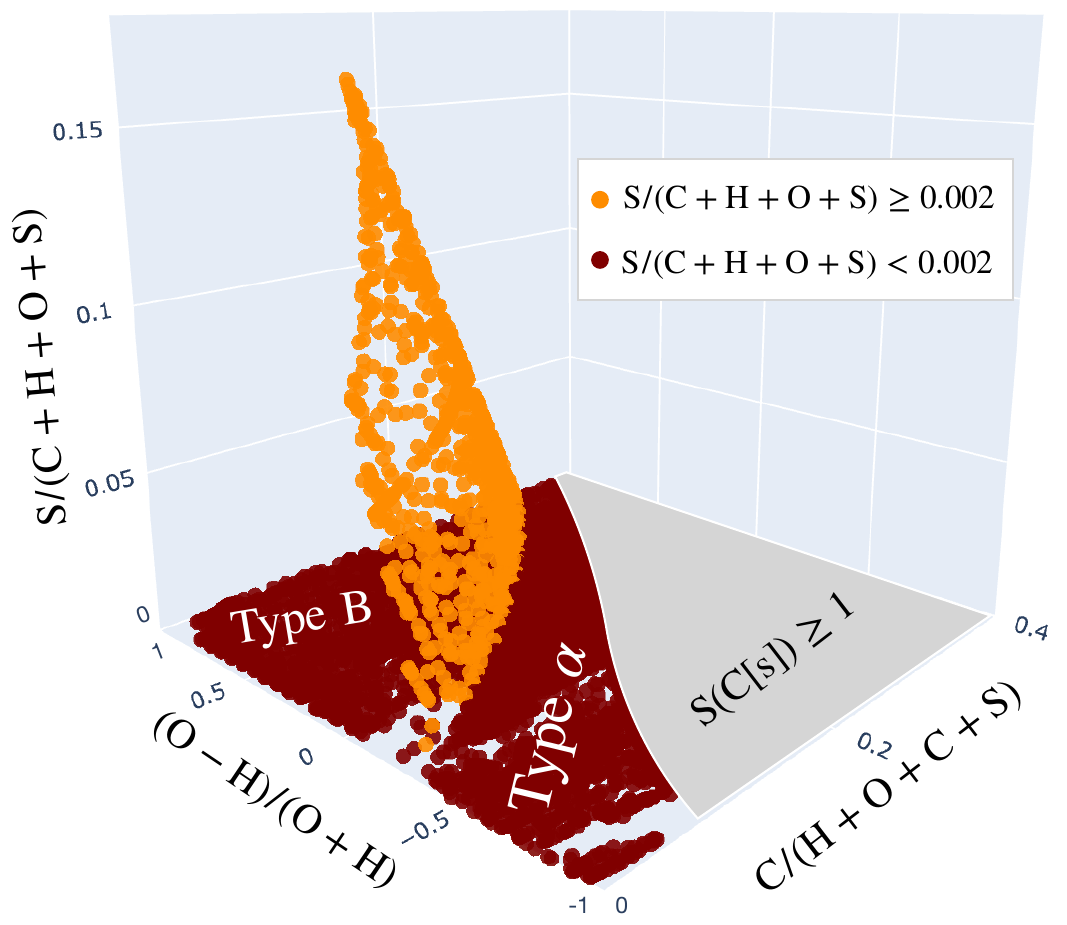}
    \caption{The atmospheric parameter space expanded by the inclusion of sulphur on a 3$^{\mathrm{rd}}$ axis. The models displayed are based on the Ryugu elemental abundance and are calculated at 900\,K. Models in bright orange have atmospheres with sulphur abundances of $S/(C+H+O+S)\geq0.002$, which classifies them as sulphur-rich. Dark red models are classified as sulphur-poor. }
    \label{fig:3Dplot}
\end{figure}

As stated by \cite{Janssen_2023}, sulphur condensation relies largely on how efficient refractory elements such as Ca, Fe, Si, Mg, and Al can be assembled in refractory elements, and whether or not some Ca and Fe is left over for sulphur to bind to. That means, if these elements are in short supply or not balanced, sulphur can also be stable as a gas in the atmosphere. This is the case when using the sets of total element abundances based on Ryugu and CI. Despite them having 5 times more iron than the Earth and CC sets of total element abundances, Ryugu and CI contain about 4 times less Ca, about 2.7 times less Si, and about 8 times less Al. Additionally, Ryugu and CI contain about 75 times more sulphur \ls{compared to Earth and CC sets of total elemental abundances}. 

When refractory condensates form from Ca, Fe, Si, Mg, and Al, an overabundance of iron is left over in models based on the Ryugu and CI elemental abundances. This results in the formation of \ce{Fe3O4}[s] (iron(II,III) oxide), which depletes oxygen from the gas phase, and in turn allows for a higher sulphur content in the gas phase. In the atmosphere, sulphur then forms \ce{SO2}, which following \cite{Janssen_2023}, would be classified as a type BC2 atmosphere (coexistence of \ce{SO2}, \ce{H2SO4}, \ce{CO2}, \ce{H2O}, and \ce{N2}). The crust chemistry is then entirely different from other models with the same CHO gas phase abundances but different S gas phase abundance. Atmospheres with $S/(C+H+O+S)\geq0.005$ at 1000\,K, $S/(C+H+O+S)\geq0.002$ at 900\,K and 800\,K, and $S/(C+H+O+S)\geq0.001$ for $T\leq700$\,K are considered sulphur-rich. Only grids based on the Ryugu and CI set of total element abundances with $T\geq600$\,K produce sulphur-rich atmospheres. We note that we only find sulphur-rich atmospheres with $(O-H)/(O+H)$ ratios above -0.3, and that the only sulphur-rich atmospheric type reached is type BC2.

An example of this is shown in Fig. \ref{fig:3Dplot}. Here, the usual CHO-parameter space is expanded by the inclusion of the atmospheric sulphur abundance on a 3$^{\mathrm{rd}}$ axis making the reason for the different classification apparent. Despite some models possessing the same CHO abundances as type B atmospheres, they are sulphur-rich motivating the necessary distinction between type B and type BC2 atmospheres. \ls{Notably, all sulphur-rich models at all temperatures are constrained to the curved and slanted vertical plane shown in Fig. \ref{fig:3Dplot}. In theory, more sulphur-rich compositions are possible (see \cite{Janssen_2023}), but not reached by the models in this work. Whether this is a physical effect or not, is a subject of future investigations.}

\subsubsection{Linked Condensates under sulphur-rich atmospheres}
\label{subsub:rich_cond}

The two Class 0 condensates identified in this work, \ce{MgSiO3}[s] and \ce{KAlSi3O8}[s] described in Sect. \ref{subsub:furth_poor_cond}, do also exist in crusts contacting type BC2 atmospheres.  \ce{MgSiO3}[s] is a Class 0 condensate at 600\,K and 700\,K, at higher temperatures its region of stability in crusts contacting atmospheres throughout the atmospheric CHO parameter space changes. From 800\,K to 1000\,K \ce{MgSiO3}[s] is a Class II condensate, existing in crusts contacting type BC2 atmospheres.  \ce{KAlSi3O8}[s] is a Class 0 condensate for $T=800$\,K to $T=1000$\,K, therefore it also exists in all crusts contacting type BC2 atmospheres. We note that there are 25 outlier crust models in which \ce{KAlSi3O8}[s] does not exist (discussed in Sect. \ref{sub:outliers}).

There are two linked sulphur condensates in crusts contacting sulphur-rich atmospheres. These are \ce{CaSO4}[s], a Class I condensate, and \ce{Na2SO4}[s], a Class II condensate. \ce{CaSO4}[s] exists in all crusts contacting BC2 atmospheres, at $T=600$\,K and $T=700$\,K. At 800\,K and 900\,K it exists in crusts contacting atmospheres with the sulphur content $S/(C+H+O+S)\geq0.6$\% and $S/(C+H+O+S)\geq6$\% respectively. Below these thresholds and at 1000\,K, \ce{CaSO4}[s] is largely replaced by FeS[s]. \ce{Na2SO4}[s] exists in crusts contacting type BC2 atmospheres from 600\,K to 800\,K. At 800\,K it only exists in crusts contacting atmospheres with sulphur abundances above the threshold of $S/(C+H+O+S)\geq0.14$\%. The existence of \ce{Na2SO4}[s] is likely a bias from the Ryugu and CI set of total element abundances, which are both low in Ca, inhibiting \ce{CaSO4}[s] formation, allowing further -\ce{SO4}[s] condensates to form. 

\ce{Fe2O3}[s] (iron(III) oxide) is a Class I condensate at 600\,K and 700\,K, existing in all crusts contacting type BC2 atmospheres.

\ce{Mn3Al2Si3O12}[s] (spessartine) is a Class I  condensate for $T\geq600$\,K, existing in every crust in phase equilibrium with type BC2 atmospheres. 

\ce{MgF2}[s] is a Class II condensate at 600\,K and is stable in all crusts contacting atmospheres with the type BC2. 

\subsection{Outliers}
\label{sub:outliers}

One group of outliers are the most reducing (hydrogen- and carbon-rich) atmospheric compositions. In the CHO-parameter space, those exist at the very bottom right, close to the $S$(C[s])=1 line. Crusts in phase equilibrium with such type $\alpha$ atmospheres have differing compositions compared to crusts in phase equilibrium with atmospheres of the same type (type A or $\alpha$). As described in Sects. \ref{subsec:sulphurpoor_atm} and \ref{subsub:iron_poor_cond}, sulphur, instead of being stable in the form of \ce{FeS}[s], occurs as \ce{MnS}[s], \ce{MgS}[s], and \ce{CaS}[s]. This is because, iron exists in its pure and unoxidised form Fe[s]. At temperatures below 1000\,K, less than 10 models per grid are such outliers. At 1000\,K a substantial triangular region containing such outliers becomes apparent (see also Fig. \ref{fig:summary}). 

We note that 25 crusts in phase equilibrium with type BC2 atmospheres are outliers with respect to the existence of \ce{KAlSi3O8}[s]. \ce{KAlSi3O8}[s] exists in all models across all grids as Class 0 condensate, except in 25 data points with the atmospheric type BC2 from 800\,K to 1000\,K.

\section{Discussion}
\label{sec:limit}
This paper is a continuation and enhancement of the investigation in \cite{oh-ls25}, where a link between the surface mineralogy and the atmospheric type was implied. 
However, each atmospheric type was only sampled by 3 to 12 models. It was therefore unclear whether transitions of the crust composition coincide with the atmospheric type transitions or not. 
Furthermore, the independence of the link between the atmospheric types and the stability of surface minerals from the relative elemental abundances of the refractory elements was beyond the investigation in \citet{oh-ls25}, which focussed on the observability of atmospheres of different atmospheric types.
Within the work at hand, we have extended the temperature range, to span from 500\,K to 1000\,K. For such planets, the assumption of an atmosphere in chemical equilibrium is more justifiable \citep[see also][]{Liggins2023}.
The assumption of thermo-chemical and phase equilibrium on which all of the models presented here are discussed in the following subsection.
While \cite{oh-ls25} sampled atmospheric types A-D and indicated the presence of a transitional regime between atmospheric types B and C, no sulphur rich atmospheres were discussed. This is a direct result from the lower temperatures, inhibiting the formation of sulphur rich atmospheric compositions.

These reasons motivate this work, for which multiple complete grids, consisting of over 5000 crust-atmosphere models for each pair of total element abundances and temperature are generated. Each of the models in such a grid is based on the same set of total elemental abundances, with varying C, H, and O abundances, effectively isolating the effect of the atmospheric composition on the mineralogical crust composition. In order to avoid biases from the refractory elemental abundances, multiple grids based on different total elemental abundances are generated at each temperature. This allows to fully sample each atmospheric type and make general statements about crustal compositions found in contact with each each atmospheric type, regardless of the overall total elemental composition. Across temperatures ranging from 500\,K to 1000\,K we generate a total of over 150000 crust-atmosphere models. In addition to the previously sampled atmospheric types A-D including the transitional region between type B and C, we also see the emergence of sulphur-rich atmospheric compositions.
Despite the possibility of various different sulphur-based atmospheric types \citep[see][]{Janssen_2023}, we only see a single one (type BC2).
Overall this work assesses the stability of 71 condensates across 5 atmospheric types.

In the following the limitations of the model assumptions and implications of the results are discussed. 

\subsection{Limitations}

A crust and atmosphere do not necessarily have to be in thermo-chemical and phase equilibrium with each other. Common processes such as e.g. volcanism, atmospheric loss, or photochemistry can disrupt or prevent the system from reaching the equilibrium state in the first place. Volcanic gases are drastically hotter than the atmosphere itself which can introduce species that should not co-exist in equilibrium at the pressure and temperature of the atmosphere. \oh{Earth's} atmosphere, according to equilibrium chemistry, should consist of \ce{O2}, \ce{N2}, \ce{CO2}, and \ce{H2O} and would be classified as type B (see also definition in Eq. \ref{eq:Types}). However, due to the presence of life, \ce{CH4} does exists in the \oh{Earth's} atmosphere, which is not possible if equilibrium chemistry is assumed. Furthermore, the lower the temperature, the longer it takes a system to reach equilibrium \oh{as reaction timescales increase with decreasing temperature} \citep[see e.g.][for a discussion on timescales]{Herbort20rocky1, Liggins2023}.
If a planet experiences moderate but consistent volcanism and has a relatively cool atmosphere ($T<700$\,K), it may never reach equilibrium \citep{Liggins2023}. Despite this fact, even if equilibrium timescales are so large that equilibrium is never reached, every system evolves towards equilibrium. 
\oh{In the case of moderate levels of disequilibrium processes, assuming equilibrium as a first order approximation will not yield drastically different results to reality.}  
This is shown in \cite{Herbort2022} on the example of \oh{the Earth's} atmosphere. Earth experiences volcanism, harbours an ecosystem, and is subject to atmospheric loss and photochemistry. Modelling \oh{Earth's} atmosphere using equilibrium chemistry yields a composition which lays within a couple of ppm of the actual composition with the exception of \ce{CH4}, showcasing  
\oh{the reasonable character of such equilibrium models.}
Similarly, Venus' atmosphere can also be modelled with \textsc{GGchem}, as demonstrated by \cite{rimmer2021hydroxide} and including the crust by \cite{Byrne24}.

The investigation of disequilibrium processes lies beyond the scope of this work. Since the term ‘disequilibrium processes’ includes a large amount of various different processes, many of which are not fully understood, the inclusion of only a couple of those would not make the model complete.

Chemical reaction networks grow exponentially with the number of included elements/species, which is at some point limited by computational power. Additionally, many reaction rates do not exist yet, leading to incomplete, potentially misleading results. Due to this, chemical reaction networks are often constrained to the CHNOS elements \citep[e.g.][]{rimmerStand2015, jordan21venusphoto, tsai2024global} which would not be suitable for the modelling of the surface composition.

As mentioned, volcanism and outgassing have been disregarded in this work. For moderately warm ($T\geq700$\,K) planets, observed during volcanically dormant periods, this is not a problem as at least the gas phase of these systems should evolve towards equilibrium \citep{Liggins2023}. Especially for forming secondary atmospheres, volcanism and outgassing are the most important processes governing the composition of the atmosphere \citep[e.g.][]{schaefer2012, volcanic_degassing2014}. 
To include these, interior models \citep[e.g][]{spaargaren2020influence, baumeisterInterior} have to be more extensively coupled to atmospheric models, which is part of future work.

The work presented here is based on \textsc{GGchem} and therefore the minimisation of Gibbs free energy of pure condensates. 
While a system of such condensates will strive towards this ideal state, it remains a question whether such state will be reached.
Here discussions on timescales for condensation, annealing, and chemisorption in order to equilibrate the condensate phase become important. 
Such discussion has been presented in Sect.~7 of \citet{Herbort20rocky1}. While the condensation\oh{, the reaction from gas phase to condensate phase, either directly or by gas-gas reactions,} is efficient due to the presence of a large surface area \oh{even at low tempartures,}  the annealing time\oh{, the restructuring of condensates to form other condensates,} rises significantly with lower temperatures, while chemisorption\oh{, the incorporation of some condensate into another existing mineral,} remains faster at similar temperatures. 
The latter is especially relevant for example for the inclusion of \ce{OH} or \ce{H2O} groups in phyllosilicates \citep{Thi2020}, allowing this process to be efficient down to temperatures even below 100\,K. 
Furthermore, minerals, which are predicted to not be thermally stable in contact with a given atmosphere should react to the stable form. 
A prominent example of a mineral formed in the mantle and brought to Earth's surface, which is decaying on short timescales under Earth's atmospheric conditions is \ce{FeS2}[s] \citep[see][and references therein]{Newman1998Pyrite, Larkin2011Pyrite}.

\oh{\citet{Charnoz2026} present a framework, which provides insight to non-equilibrium condensation in a protoplanetary disk. 
In their model, they assume a gas phase in chemical equilibrium at all temperatures as a basis for their condensation.
The condensation kinetics are parameterised by gas-gas reactions for the initial formation of condensates and gas-grain reactions, which transforms one mineral into another.
\citet{Charnoz2026} investigate pressures of $10^{-9}\,$bar to $10^{-2}\,$bar and cooling timescales from 0.01\,yr to 1000\,yr.
For models towards the high end of their parameter space, they suggest good agreement between equilibrium and kinetic condensation, if the temperature is high enough ($T\geq 800$\,K).
Further extrapolating these models to the surface mineralogy of rocky planets introduces changes in pressure by orders of magnitudes and necessitates investigations over geological timescales on the order of Gyr. In the model presented here, we focus on atmospheric pressures of 1\,bar, of which only a small subset is hydrogen dominated.
Adapting this model to different atmospheric compositions, higher pressures, and longer timescales should increase the parameter space, in which equilibrium condensation is applicable, as all of these changes should decrease the  gas-grain reaction rates discussed in \citet{Charnoz2026}.
Extending such a condensation kinetics model to planetary environments is beyond the scope of this paper. 
}

Furthermore, the current implementation of \textsc{GGchem} does not include effects such as solid solutions and the formation of partial melts. 
Further tools such as \textsc{Perple\_X} \citep{connolly09}, \textsc{ExoPlex} \citep{Exoplex}, or \textsc{MAGEMin} \citet{Riel2022} could be used to investigate changes induced by such effects.
\oh{These can shine light onto the stability of minerals found in this study, both on the surface and in the interior, for example in case of non-stagnant lid planets.
Here, the results from this work can be used as a boundary conditions for the surface mineralogy.}

The models investigated in this work only include the solid surface and the atmospheric layer in direct contact with the surface. However, transit observations are only able to probe the upper layers of any given atmosphere, where pressures are relatively low \citep[e.g][]{madhusudhan2009temperature}. This leads to the possible scenario where atmospheres which are chemically distinct at the base layer, lose their chemical uniqueness because of chemical evolution with height  \citep[e.g. condensation of cloud particles, see also][]{Mbarek16clouds, loftus19clouds,Herbort2022}. This can then blur the sharp transitions in crust composition seen in this work in real life observations. Previous works show that if a  $p,T$-profile is assumed atmospheric compositions can in fact converge or diverge along the  $p,T$-profile. However this effect is less pronounced at higher temperatures ($T\geq600$\,K), and does not change the atmospheres type, except for atmospheres with high nitrogen content which can lead to atmospheres changing types from C to A with altitude \citep{oh-ls25}. Further study is needed, possibly a similar grid study with the inclusion of a $p,T$-profile for every grid point. Since the shape of the $p,T$-profile is largely governed by the incoming stellar flux, future work could incorporate stellar flux into $p,T$-profile calculation \citep[e.g. with \textsc{ARCiS}][]{min2020arcis}.

The modelling in this work neglects the shape of planets, which are inherently 3D objects. For example, the day and night side of a planet have different temperatures, which can cause significantly different patterns in atmospheric and surface composition as well as cloud coverage \citep[e.g.][]{Kite2016, turbet2021day, Powell2024}. Extreme cases would be tidally locked planets, with one side in eternal day and the other side in eternal darkness. This can lead to drastic temperature differences, which would change both the crusts and atmospheres composition. A similar effect is seen at the poles, where the incident angle of stellar radiation is shallower, causing lower temperatures. For the results of this work to apply, geological timescales should be much longer than the rotation period of the planet.

The overall element availability does play a big role in the final composition of a planet. In this work different sets of total element abundances lists were used to prevent biases. Nonetheless, all sets of total element abundances are based on bodies of the solar system, which in itself is a bias towards the solar systems composition. To generate models with crust compositions that are as accurate and complete as possible, 18 input elements were used. Elemental abundances found in stars or in protoplanetary disks could inform exoplanet modelling, but currently focus on mostly volatiles \citep[e.g.][]{COS_Disk}. Current modelling of exoplanets compositions based on stellar compositions \citep{spaarWangbulkcomp,spaargaren25disk}, and future disk observations might provide more complete elemental abundance measurements to inform terrestrial exoplanet (crust and bulk composition) modelling.

If an atmosphere consists of more than around 1\% sulphur content, an atmosphere could potentially fall into an atmospheric subtype defined by the existence of sulphur species. This changes the crusts composition compared with crusts contacting atmospheres with the same x and y coordinates in the CHO parameter space. This work offers some constraints on crusts in phase equilibrium with type BC2 atmospheres. However, other sulphur rich atmospheric types as described in \cite{Janssen_2023} require further investigation of their presence at these temperatures requires additional work. 
This work shows that based on our models up to 1000\,K, none of the other subtypes exist.

\begin{figure*}
    \centering
    \includegraphics[width=1\linewidth]{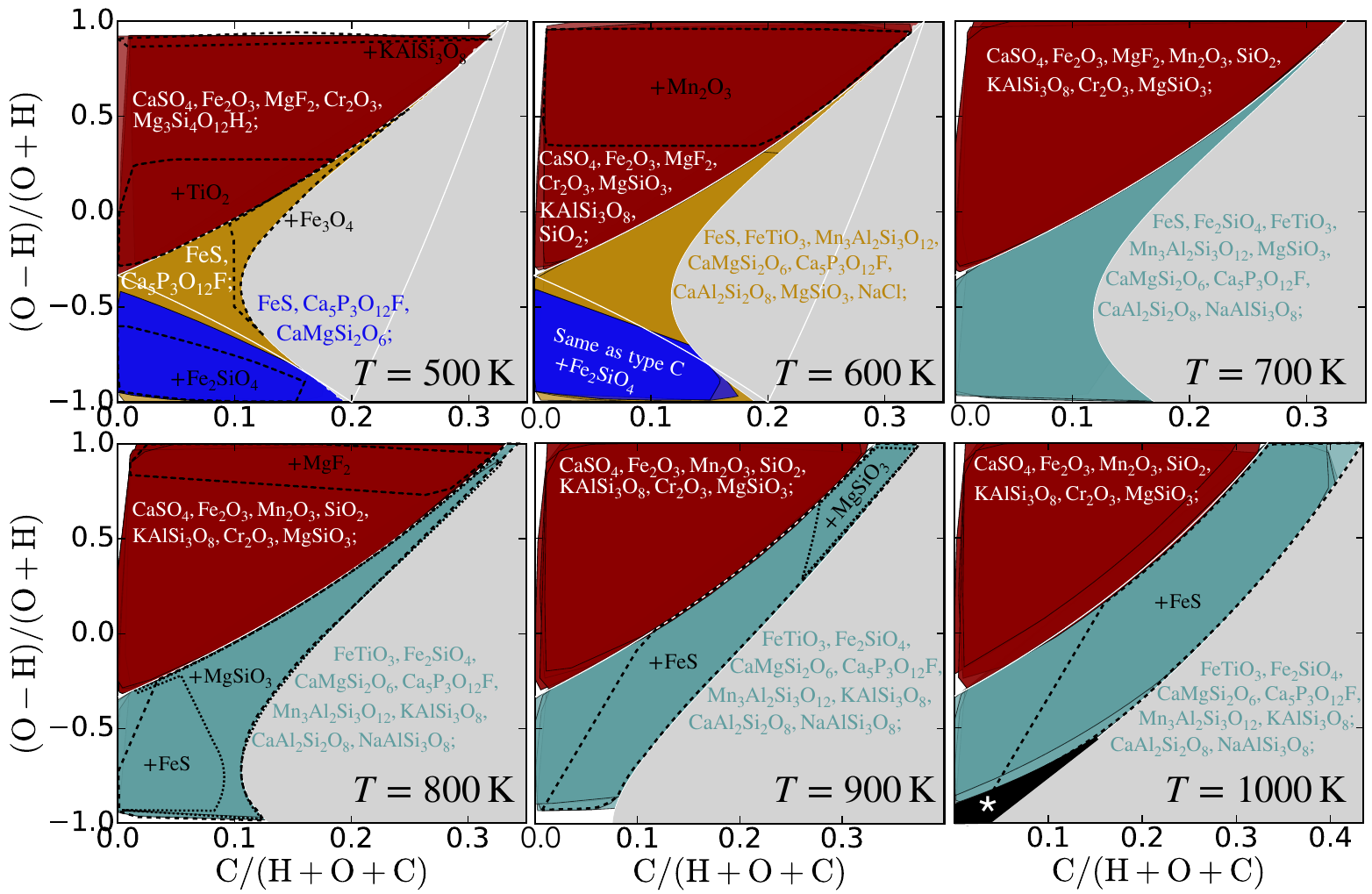}
    \caption{All condensates linked to sulphur-poor atmospheric types A, B, C, and $\alpha$ at all temperatures investigated in this work. Condensates found in all crusts in phase equilibrium with all compositions in an atmospheric type are indicated with white or coloured text. Regions in the parameter space where condensates are stable in crusts in phase equilibrium with a subset of atmospheric compositions of an atmospheric type are indicated with black dashed outlines.  Gray surfaces indicate where graphite is supersaturated, white areas indicate missing data. Note that condensates in crusts in phase equilibrium with type BC2 and  the transitional regions between types B and C/$\alpha$ are excluded from this figure. For a full description see Sect. \ref{sec:conclusion} and Appendix \ref{app:links}. $\ast$: The black triangular area in the parameter space for 1000\,K encompasses heavily reducing atmospheres where crust compositions differ. Due to heavily reducing conditions, iron exists as unoxidised Fe[s], causing sulphur to exist as MnS[s], MgS[s], and CaS[s].}
    \label{fig:summary}
\end{figure*}

\subsection{Implications}

The condensates described in this work in Sect. \ref{sec:results} (see also Fig. \ref{fig:summary} and the Tables in Appendix \ref{app:links}), are linked to atmospheric type. If a future observation constrains a terrestrial exoplanets atmosphere to an atmospheric type \oh{and a certain surface temperature}, by for example observing features of the defining gas phase molecules (listed in Equations \ref{eq:Types} and \ref{eq:TypeshighT}, model spectra in \cite{oh-ls25}) or constraints on the abundances of the elements C, H, and O, this work can in principle offer constraints on the surface`s composition. 

Changes in pressures and temperatures along a $p,T$-profile,  can lead to similarly sharp transitions of condensate species as seen in crusts in phase equilibrium with atmospheres in the CHO-parameter space in this work. \cite{Byrne24} demonstrate this on the example of Venus, tracking condensate species along the $p,T$-profile of Venus. This motivates further investigation of pressure space, which could be done by repeating this works grid study at different pressures.

With the findings of this work the highly degenerate question of constraining compositions of terrestrial exoplanet surfaces is a step closer to being answered. 
\oh{Besides observationally determining the atmospheric composition to an atmospheric type, it is also necessary to know the planetary surface temperature and pressure sufficiently precisely in order to accurately assess the presence of specific surface minerals.}
To apply the findings of this work, new observational techniques are not required, rather, existing and constantly improving observational techniques such as transit spectroscopy can be employed. The transitions of surface mineralogy are sudden in the CHO-parameter space and often stable over a range of temperatures. The fact that the crust composition can be linked to the corresponding atmospheric type means that future observations do not have to be precise enough to constrain the exact CHO elemental abundances of the atmosphere for inferring the crustal composition.  Instead, constraining the atmosphere to a specific type allows to put constraints on the surface mineralogy.

\section{Summary}
\label{sec:conclusion}

This work has investigated the link between atmospheric and surface compositions for rocky planets with 1\,bar atmospheres across temperatures from 500\,K to 1000\,K.  It continues the investigation done in \cite{oh-ls25}, and better resolves the ambiguous link, thanks to the much finer sampling of the CHO-atmospheric parameter space. The model grids are based on Earth, CC, Ryugu, CI, BSE, and MORB total element abundances to prevent biases. 
Based on the results described in sections \ref{subsec:sulphurpoor_atm} and \ref{subsec:sulphurrich_atm}, it is possible to predict a range of crust condensates per atmospheric type and per temperature. The abundances of  condensate species are displayed in Fig.~\ref{fig:TypeOverview} for a few representative models.

\oh{The major findings of this paper can be summarised as follows:
\begin{itemize}
    \item The atmospheric compositions reach distinct types, defined especially by the presence of \ce{O2}, \ce{SO2}, \ce{H2}, and the co-existence of \ce{CH4} and \ce{CO2}.
    \item While in theory the 3D parameter space of C, S, and ${O-H/O+H}$ could be populated, we find only two distinct planes occurring (see Fig.~\ref{fig:3Dplot}).
    \item Many transitions in condensates follow the changes in atmospheric types directly and suggest that the thermal stability of some minerals can in principle be constrained by observations of the atmosphere (see Fig.~\ref{fig:summary}).
    \item The oxidisation state of iron follows distinct changes, reflecting the atmospheric types (see Fig.~\ref{fig:iron_poor}).
\end{itemize}}

\oh{A more detailed overview of the different minerals stable for specific atmospheric types at a given temperature is provided in the following.}
At 500\,K, atmospheric types A, B, C, and the transitional regime exist. At 600\,K, the same types with the addition of type BC2 exist. From 700\,K and onwards to 1000\,K, atmospheric types A and C merge into type $\alpha$, type B and the transitional regime stay the same compared to lower temperatures, and higher and higher sulphur abundances are possible in atmospheric compositions in type BC2 as the temperature rises.

\paragraph{500\,K:}\begin{description}
    \item[\textbf{Type A:}]  \ce{FeS}[s],  \ce{CaMgSi2O6}[s], \ce{Ca5P3O12F}[s];\\
    In a subset of type A: \ce{Fe2SiO4}[s];
    \item[\textbf{Type B:}]  \ce{CaSO4}[s], \ce{Fe2O3}[s], \ce{MgF2}[s], \ce{Cr2O3}[s], \ce{Mg3Si4O12H2}[s];\\ In a subset of type B:  \ce{KAlSi3O8}[s], \ce{TiO2}[s];
    \item[\textbf{Type C:}]  \ce{FeS}[s], \ce{Ca5P3O12F}[s];\\ In a Subset of type C: \ce{Fe3O4}[s];
    \item[\textbf{Type BC2:}] Non-existent
    \item[\textbf{Transitional regime:}] \ce{FeS2}[s], \ce{FeTiO3}[s], \ce{Mn3Al2Si3O12}[s];
\end{description}

\paragraph{600\,K:}\begin{description}
    \item[\textbf{Type A:}]  \ce{FeS}[s],  \ce{FeTiO3}[s],  \ce{Mn3Al2Si3O12}[s], \ce{Fe2SiO4}[s], \ce{CaMgSi2O6}[s], \ce{Ca5P3O12F}[s],  \ce{CaAl2Si2O8}[s], \ce{MgSiO3}[s], \ce{NaCl}[s];
    \item[\textbf{Type B:}]  \ce{CaSO4}[s], \ce{Fe2O3}[s], \ce{MgF2}[s], \ce{MgSiO3}[s], \ce{SiO2}[s],  \ce{KAlSi3O8}[s], \ce{Cr2O3}[s];\\ In a subset of type B: \ce{Mn2O3}[s];
    \item[\textbf{Type C:}] \ce{FeS}[s], \ce{FeTiO3}[s], \ce{Mn3Al2Si3O12}[s], \ce{CaMgSi2O6}[s], \ce{Ca5P3O12F}[s],  \ce{CaAl2Si2O8}[s], \ce{MgSiO3}[s], \ce{NaCl}[s];
    \item[\textbf{Type BC2:}]  \ce{CaSO4}[s],  \ce{Na2SO4}[s], \ce{Fe2O3}[s], \ce{MgF2}[s], \ce{Mn3Al2Si3O12}[s], \ce{MgSiO3}[s];
    \item[\textbf{Transitional regime:}] \ce{FeS2}[s], \ce{Fe3O4}[s], \ce{CaTiSiO5}[s], \ce{MgSiO3}[s];
\end{description}

\paragraph{700\,K:}\begin{description}
    \item[\textbf{Type $\alpha$:}]  \ce{FeS}[s],  \ce{FeTiO3}[s],  \ce{Mn3Al2Si3O12}[s], \ce{Fe2SiO4}[s], \ce{CaMgSi2O6}[s], \ce{Ca5P3O12F}[s],  \ce{CaAl2Si2O8}[s], \ce{NaAlSi3O8}[s], \ce{MgSiO3}[s];
    \item[\textbf{Type B:}] \ce{CaSO4}[s], \ce{Fe2O3}[s], \ce{MgF2}[s], \ce{Mn2O3}[s], \ce{SiO2}[s],  \ce{KAlSi3O8}[s], \ce{Cr2O3}[s], \ce{MgSiO3}[s];
    \item[\textbf{Type BC2:}] \ce{CaSO4}[s],  \ce{Na2SO4}[s], \ce{Fe2O3}[s], \ce{Mn3Al2Si3O12}[s], \ce{MgSiO3}[s];
    \item[\textbf{Transitional regime:}] \ce{FeS2}[s], \ce{Fe3O4}[s], \ce{CaTiSiO5}[s], \ce{MgSiO3}[s];
\end{description}

\paragraph{800\,K:}
\begin{description}
    \item[\textbf{Type $\alpha$:}] \ce{FeTiO3}[s], \ce{Fe2SiO4}[s],  \ce{CaMgSi2O6}[s], \ce{Ca5P3O12F}[s], \ce{Mn3Al2Si3O12}[s],  \ce{CaAl2Si2O8}[s], \ce{NaAlSi3O8}[s], \ce{KAlSi3O8}[s];\\ In a subset of type $\alpha$: \ce{FeS}[s], \ce{MgSiO3}[s];
    \item[\textbf{Type B:}]  \ce{CaSO4}[s], \ce{Fe2O3}[s], \ce{Mn2O3}[s], \ce{SiO2}[s], \ce{KAlSi3O8}[s], \ce{Cr2O3}[s], \ce{MgSiO3}[s];\\ In a subset of type B: \ce{MgF2}[s];
    \item[\textbf{Type BC2:}] \ce{Mn3Al2Si3O12}[s], \ce{MgSiO3}[s], \ce{KAlSi3O8}[s];\\ In a subset of type BC2: \ce{CaSO4}[s],  \ce{Na2SO4}[s];
    \item[\textbf{Transitional regime:}] \ce{FeS2}[s], \ce{Fe3O4}[s], \ce{KAlSi3O8}[s];
\end{description}

\paragraph{900\,K:}
\begin{description}
    \item[\textbf{Type $\alpha$:}]\ce{FeTiO3}[s], \ce{Fe2SiO4}[s],  \ce{CaMgSi2O6}[s], \ce{Ca5P3O12F}[s], \ce{Mn3Al2Si3O12}[s],  \ce{CaAl2Si2O8}[s], \ce{NaAlSi3O8}[s], \ce{KAlSi3O8}[s];\\ In a subset of type $\alpha$: \ce{FeS}[s], \ce{MgSiO3}[s];
    \item[\textbf{Type B:}]  \ce{CaSO4}[s], \ce{Fe2O3}[s], \ce{Mn2O3}[s], \ce{SiO2}[s], \ce{KAlSi3O8}[s], \ce{Cr2O3}[s], \ce{MgSiO3}[s];
    \item[\textbf{Type BC2:}] \ce{Mn3Al2Si3O12}[s], \ce{MgSiO3}[s], \ce{KAlSi3O8}[s];\\ In a subset of BC2: \ce{CaSO4}[s];
    \item[\textbf{Transitional regime:}]  \ce{Fe3O4}[s], \ce{KAlSi3O8}[s];
\end{description}

\paragraph{1000\,K:}
\begin{description}
    \item[\textbf{Type $\alpha$:}] \ce{FeTiO3}[s], \ce{Fe2SiO4}[s],  \ce{CaMgSi2O6}[s], \ce{Ca5P3O12F}[s], \ce{Mn3Al2Si3O12}[s],  \ce{CaAl2Si2O8}[s], \ce{NaAlSi3O8}[s], \ce{KAlSi3O8}[s];\\ In a subset of type $\alpha$: \ce{FeS}[s];
    \item[\textbf{Type B:}]   \ce{CaSO4}[s], \ce{Fe2O3}[s], \ce{Mn2O3}[s], \ce{SiO2}[s], \ce{KAlSi3O8}[s], \ce{Cr2O3}[s], \ce{MgSiO3}[s];
    \item[\textbf{Type BC2:}] \ce{Mn3Al2Si3O12}[s], \ce{MgSiO3}[s], \ce{KAlSi3O8}[s];
    \item[\textbf{Transitional regime:}]  \ce{Fe3O4}[s], \ce{KAlSi3O8}[s];
\end{description}

These are the crustal compositions which can be expected on terrestrial exoplanet surfaces given atmospheric type, with the respective temperatures, and 1\,bar of surface pressure. A visualisation of this list is provided in Fig. \ref{fig:summary}, excluding type BC2 and the transitional regions between types B and C/$\alpha$.

\begin{acknowledgements}
The authors thank R.J. Spaargaren and L. M\"uller for their valuable discussions on (exoplanet) mineralogy. LS acknowledges funding by the Research Council of Norway (RCN), through its Centres of Excellence funding scheme, projects number 332523 (PHAB, Centre for Planetary Habitability), and the researcher project number 358248 (GREP, Galactic Recipe for Exo-Planets).
OH acknowledges financial support by FFG under grant number 62545255. 
\end{acknowledgements}

\bibliography{library.bib}
\onecolumn
\appendix

\section{set of total element abundances}
\label{app:elemental-inputs}

\begin{table*}[htp]
\caption[]{\label{tab:abundances1}Elemental mass fractions of the sets of element abundances used in this work.}
\begin{center}
\begin{tabular}{l|cccccc}
\hline \hline
        & Earth    & CC & CI & Ryugu & MORB & BSE \\
        &    1     &  2 &  3 &    4  &   5  & 2   \\
\hline
H        & 0.3341 & 0.045  & 1.992  & 1.034159      &  0.023   & 0.006\\
C        & 1.957  & 0.199  & 3.52   & 4.514885      &  0.019   & 0.006\\
O        & 50.1   & 47.2   & 46.42  & 39.694165    &  44.5    & 44.42\\
\hline
\hline
C+H+O  & 52.3911& 47.444 & 51.932 & 45.243209     &  44.542  & 44.432\\
\hline
\hline
N        & 0.0253 & 0.006  & 0.289  & 0.146115      &  5.5E-05 & 8.8E-05\\
F        & 0.0482 & 0.053  & 0.0059  & 0.0241 $\ast$&  0.017   & 0.002\\
Na       & 2.145  & 2.36  & 0.505  & 0.612898       &  2.012   & 0.29 \\
Mg       & 1.999  & 2.20  & 9.790  & 11.126626      &  4.735   & 22.01 \\
Al       & 7.234  & 7.96  & 0.8680  & 0.917588      &  8.199   & 2.12 \\
Si       & 26.17  & 28.8 & 10.820 & 12.613700       &  23.62   & 21.61 \\
P        & 0.0691 & 0.076  & 0.098  & 0.130840      &  0.057   & 0.008\\
S        & 0.0636 & 0.07 & 5.411 & 5.729325         &  0.11    &  0.027 \\
Cl       & 0.0427 & 0.047 & 0.071  & 0.077981       &  0.014   & 0.004\\
K        & 1.945  & 2.14 & 0.055  & 0.052457        &  0.152   & 0.02\\
Ca       & 3.499  & 3.85 & 0.933  & 1.397637        &  8.239   & 2.46\\
Ti       & 0.3644 & 0.401 & 0.046  & 0.047633       &  0.851   & 0.12\\
Cr       & 0.0118 & 0.013 & 0.268  & 0.269418       &  0.033   & 0.29 \\
Mn       & 0.0654 & 0.072  & 0.195  & 0.250123      &  0.132   & 0.11\\
Fe       & 3.926  & 4.32  & 18.710 & 20.067360      &  7.278   & 6.27\\
\hline  
  
\end{tabular}\\  
\end{center}
Notes: The mass fraction values of H, C, and O are the original values of the listed set of elemental abundances. These will be different in each individual model making up a crust-atmosphere model grid. However the sum of $C+H+O$ (indicated in row 4) is kept constant in each model in the respective crust-atmosphere model grid. All remaining elemental abundances are constant and the same in each individual model across a model grid. \\
$\ast$: Value substituted as described in Eq. \ref{eq:Fryugu}. \\
1: \cite{Herbort2022}
2: \cite{schaefer2012}
3: \cite{loddersCI}
4: \cite{yokoyama2025elementalabundancesryuguassessment}
5: \cite{arevaloMORB}
\end{table*}

\section{Crust condensate abundances}
\label{app:crustabund}

\begin{sidewaysfigure*}
\centering
    \includegraphics[width=1\textwidth]{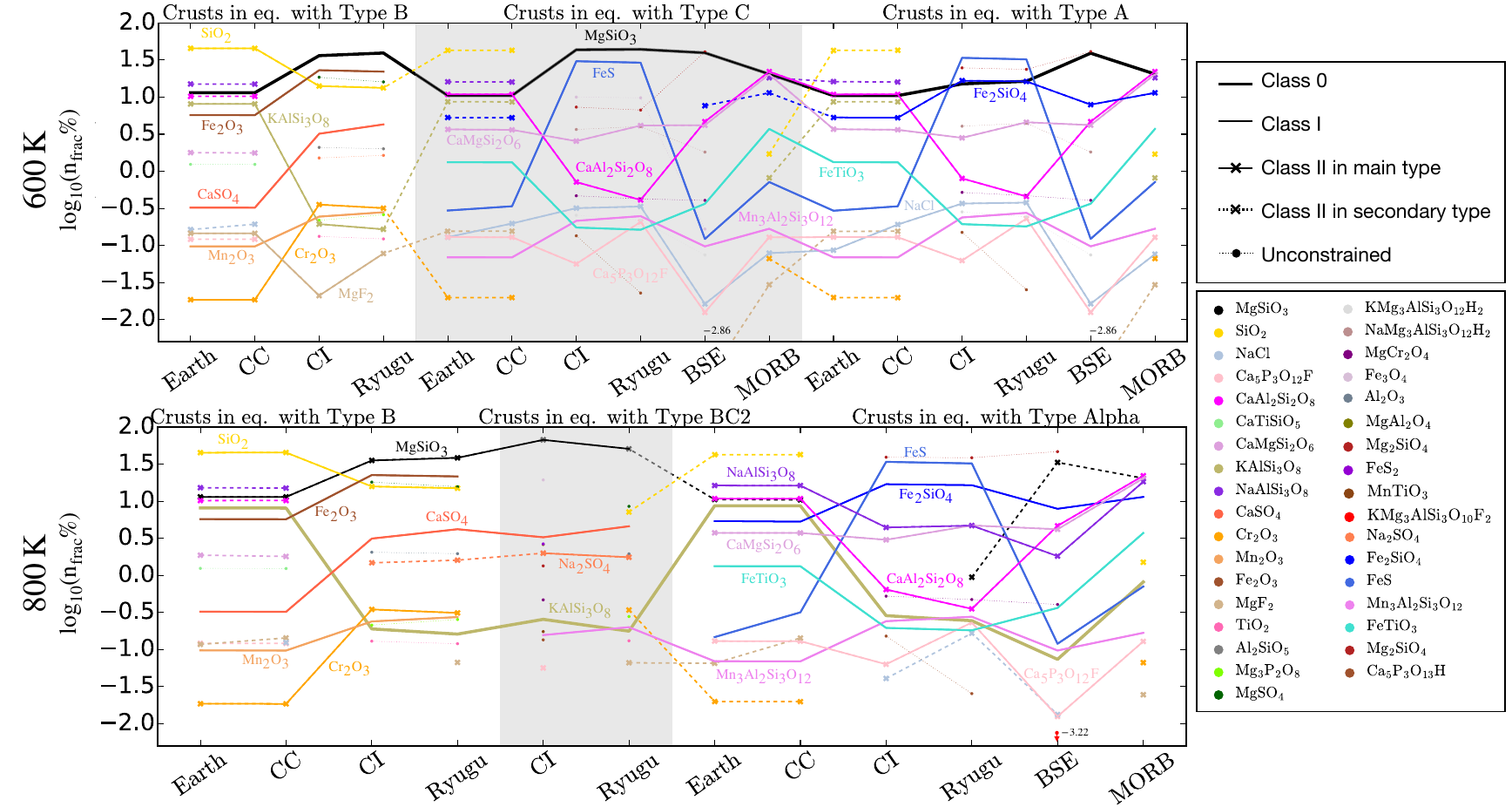}
    \caption{The abundances of crust condensates are shown in number fractions \% for specified total elemental abundances. All crust compositions are in phase equilibrium with the specified atmospheric type. Each crust composition displayed is the closest model to specified parameter space coordinates (600\,K: Type A, C/(O+H+C)=0.05, (O-H)/(O+H)=-0.75; Type C, C/(O+H+C)=0.08, (O-H)/(O+H)=-1/3; Type B, C/(O+H+C)=0.1, (O-H)/(O+H)=0.5; 800\,K: Type Alpha, C/(O+H+C)=0.08, (O-H)/(O+H)=-0.7; Type B, C/(O+H+C)=0.1, (O-H)/(O+H)=0.5; Type BC2, C/(O+H+C)=0.1, (O-H)/(O+H)=0.5, S/(C+H+O+S)=0.06, with exclusion of all sulphur-poor models) respectively and is representative of crusts in phase equilibrium with the entire atmospheric type. Abundances of individual condensates may change in crusts found in phase equilibrium throughout the respective type, however, the set of Class 0, Class I, and Class II (in main type) condensates do not change (except in rare cases see e.g. \ce{FeS}[s] at $T\geq800$\,K in Table~\ref{tab:crustcond2}. All constrained condensates (Class 0-II) are labelled in the plot and the legend, while unconstrained (Class III) condensates are only labelled in the legend.}
    \label{fig:TypeOverview}
\end{sidewaysfigure*}

\newpage
\section{Linked condensates}
\label{app:links}

\begin{table*}[!h]
\caption[]{\label{tab:crustcond1}Overview of classification of different minerals for temperatures of 500\,K, 600\,K, and 700\,K.}
\centering
\resizebox{1\textwidth}{!}{
\begin{tabular}{c | ccc|ccc|ccc}
    \hline \hline
    & \multicolumn{3}{c}{500K} &  \multicolumn{3}{c}{600K} &  \multicolumn{3}{c}{700K}\\        
    & Class 0 & Class I & Class II & Class 0 & Class I & Class II  & Class 0 & Class I & Class II \\                   
    \hline
    \ce{MgSiO3}[s] &&&&X&&&X&&\\
    \ce{CaSO4}[s] && B &&& B, BC2 &&& B, BC2 &\\
    \ce{Na2SO4}[s] &&&&&& BC2$_\mathrm{B}$ &&& BC2$_\mathrm{B}$\\
    \ce{Fe2O3}[s] && B &&& B, BC2 &&& B, BC2 &\\
    \ce{MgF2}[s] &&& B$_\mathrm{A\star}$&&& B, BC2$_\mathrm{A,C}$ &&& B$_\alpha$\\
    \ce{TiO2}[s] && B$\ast$ &&&&&&&\\
    \ce{Mg3Si4O12H2}[s] && & B$_\mathrm{A, C}$&&&&&&\\
    \ce{Mn2O3}[s] &&&&& B$\dagger$ &&& B \\
    \ce{FeS2}[s] && T &&& T &&& T &\\
    \ce{FeTiO3}[s] &&& T$_\mathrm{A,C}$ &&A,C&&& $\alpha$\\
    \ce{FeS}[s] && A, C &&& A, C&&& $\alpha$ &\\
    \ce{SiO2}[s] &&&&&& B$_\mathrm{A,C}$ &&&B$_\alpha$\\
    \ce{KAlSi3O8}[s] &&& B$_\mathrm{A, C}\otimes$&&& B$_\mathrm{A,C}$ &&& B$_\alpha$ \\
    \ce{Cr2O3}[s] &&& B$_{\mathrm{A,C}}$&&& B$_{\mathrm{A,C}}$ &&& B$_\alpha$\\
    \ce{Fe3O4}[s] &&C$\bigtriangleup$&&&&T$_\mathrm{C}$&& T \\
    \ce{NaCl}[s] &&& &&& A,C$_\mathrm{B}$&&&\\
    \ce{Mn3Al2Si3O12}[s] &&& T$_{\mathrm{A,C}}$&& A, C, BC2 &&& $\alpha$, BC2 &\\
    \ce{Fe2SiO4}[s] &&& A$_\mathrm{C}\bigtriangledown$&&& A$_\mathrm{C}$&& $\alpha$\\
    \ce{CaTiSiO5}[s] &&&&&& T$_\mathrm{B}$&&& T$_\mathrm{B}$\\
    \ce{CaMgSi2O6}[s] &&& A$_{\mathrm{C, B}}$&&&A,C$_{\mathrm{B}}$&&& $\alpha_\mathrm{B}$\\
    \ce{Ca5P3O12F}[s] &&& A,C$_{\mathrm{B}}$&&& A,C$_{\mathrm{B}}$ &&& $\alpha_\mathrm{B}$\\
    \ce{CaAl2Si2O8}[s] &&&&&& A,C$_{\mathrm{B}}$ &&& $\alpha_\mathrm{B}$\\
    \ce{NaAlSi3O8}[s] &&&&&&  &&& $\alpha_\mathrm{B}$\\
    \hline
\end{tabular}
}
\begin{flushleft}
Notes: Existence of a crust species at a certain temperature is indicated with which atmospheric type(s) the condensate exists. Lowercase text in Class II indicates under which atmospheric type the condensate has also appeared depending on the elemental input composition. \\
The transitional class of atmospheric compositions between types C and B (500\,K, 600\,K), and types $\alpha$ and B (700\,K, 800\,K, 900\,K, 1000\,K), is referred to as T.\\
$\star$: \ce{MgF2}[s] can also exist in crusts in phase equilibrium with atmospheres in a subregion of type A below  $(O-H)/(O+H)=-0.9$ depending on the input composition.\\
$\ast$: \ce{TiO2}[s] is only present below $(O-H)/(O+H) = 0.3$.\\
$\bigtriangleup$: \ce{Fe3O4}[s] is present in a complex subset of C.\\
$\bigtriangledown$: \ce{Fe2SiO4}[s] is only present in crusts in  phase equilibrium with a subset of type A and is also present in crusts where graphite is stable in crusts contacting type C atmospheres.\\
$\otimes$: \ce{KAlSi3O8}[s] is only present in all grids in crusts contacting type B atmospheres where $(O-H)/(O+H) \geq 0.9$. \\
$\dagger$: \ce{Mn2O3}[s] is only present in crusts contacting atmospheres with $(O-H)/(O+H) \geq 0.3$. Below that, Earth and CC input abundances yield \ce{Ca2MnAl2Si3O13H}[s] as the dominant Mn condensate.
\end{flushleft}
\end{table*}

\begin{table*}[]
\caption[]{\label{tab:crustcond2}Like Table~\ref{tab:crustcond1}, but for temperatures of 800\,K, 900\,K, and 1000\,K.}
\centering
\resizebox{1\textwidth}{!}{
\begin{tabular}{c | ccc|ccc|ccc}
    \hline \hline
    & \multicolumn{3}{c}{800K} & \multicolumn{3}{c}{900K} &  \multicolumn{3}{c}{1000K}\\        
    & Class 0 & Class I & Class II & Class 0 & Class I & Class II  & Class 0 & Class I & Class II \\                   
    \hline
    \ce{MgSiO3}[s]        && &B, BC2, $\alpha \odot$ &&& B, BC2, $\alpha \odot$  &&& B, BC2$_\alpha$ \\
    \ce{CaSO4}[s]         && B, BC2$\otimes$ &&& B, BC2$\bigtriangleup$ &&& B \\ 
    \ce{Na2SO4}[s]        &&& BC2$_\mathrm{B}\bigtriangledown$ &&& \\
    \ce{Fe2O3}[s]         && B &&& B &&& B \\
    \ce{MgF2}[s]          &&& B$_\alpha\ast$ &&&&&\\
    \ce{Mn2O3}[s]         && B &&& B &&& B \\
    \ce{FeS2}[s]          && T &&&&&&\\
    \ce{FeTiO3}[s]        && $\alpha$ &&& $\alpha$ &&& $\alpha \times$ \\  
    \ce{FeS}[s]           && $\alpha\ddagger$ &&& $\alpha\ddagger$ &&& $\alpha\ddagger$ \\
    \ce{SiO2}[s]          &&& B$_\alpha$ &&& B$_\alpha$ &&& B$_\alpha$\\
    \ce{KAlSi3O8}[s]      & X$\dagger$ &&& X$\dagger$ &&& X$\dagger$ &\\    
    \ce{Cr2O3}[s]         &&& B$_\alpha$ &&& B$_\alpha$ &&&  B$_\alpha$\\  
    \ce{Fe3O4}[s]         && T &&& T &&& T \\ 
    \ce{Mn3Al2Si3O12}[s]  && $\alpha$, BC2 &&& $\alpha$, BC2 &&& $\alpha \times$, BC2\\       
    \ce{Fe2SiO4}[s]       && $\alpha$ &&& $\alpha$ &&& $\alpha \times$\\   
    \ce{CaMgSi2O6}[s]     &&& $\alpha_\mathrm{B}$ &&& $\alpha_\mathrm{B}$ &&& $\alpha_\mathrm{B}$\\     
    \ce{Ca5P3O12F}[s]     &&& $\alpha_\mathrm{B}$ &&& $\alpha_\mathrm{B}$  &&& $\alpha_\mathrm{B}$\\     
    \ce{CaAl2Si2O8}[s]    &&& $\alpha_\mathrm{B}$ &&& $\alpha_\mathrm{B}$ &&& $\alpha_\mathrm{B}$ \\     
    \ce{NaAlSi3O8}[s]     &&& $\alpha_\mathrm{B}$ &&&  $\alpha_\mathrm{B}$ &&& $\alpha_\mathrm{B}$\\     
    \hline
\end{tabular}
}
\begin{flushleft}
Notes: Existence of a crust species at a certain temperature is indicated with which atmospheric type(s) the condensate exists. Lowercase text in Class II indicates under which atmospheric type the condensate has also appeared depending on the elemental input composition. \\
The transitional class of atmospheric compositions between types C and B (500\,K, 600\,K), and types $\alpha$ and B (700\,K, 800\,K, 900\,K, 1000\,K), is referred to as T.\\
$\otimes$: \ce{CaSO4}[s] only exists in crusts contacting type BC2 atmospheres, if $S/(C+H+O+S)$ is greater than 0.6\%.\\
$\odot$: \ce{MgSiO3}[s] is only present in a subregion of crusts contacting type $\alpha$\\
$\bigtriangleup$: \ce{CaSO4}[s] only exists in crusts contacting type BC2 atmospheres, if $S/(C+H+O+S)$ is greater than 6\%.\\
$\bigtriangledown$: \ce{Na2SO4}[s] only exists in crusts contacting BC2 atmospheres, if  $S/(C+H+O+S) \geq 0.14$. \\
$\ast$: \ce{MgF2}[s] is present above $(O-H)/(O+H) = 0.9$.\\
$\dagger$: \ce{KAlSi3O8}[s] is not present in 25 out of 61701 total model crust-atmosphere pairs from 800\,K to 1000\,K. These outliers have the atmospheric type BC2.\\
$\ddagger$: In model grids based on the Earth set of total element abundances, \ce{FeS}[s] only exists in a subset of type $\alpha$ contacting crusts.\\
$\times$: These condensates do not exist in crusts contacting the most reducing part of type $\alpha$
\end{flushleft}
\end{table*}

\label{LastPage}
\end{document}